\documentstyle[11pt]{article}

\newcommand{\beqa}{\begin{eqnarray}} 
\newcommand{\eeqa}{\end{eqnarray}} 
\newcommand{\beq}{\begin{equation}} 
\newcommand{\eeq}{\end{equation}}

\begin{document} 
 
\title{Superexchange and lattice distortions in the spin--Peierls 
system CuGeO$_{3}$. } 
\author{Wiebe Geertsma$^{1,2)}$ and D. Khomskii$^{2)}$, \\ 
1) UFES-CCE/Depto de Fisica,\\ 
Fernando Ferrari s/n, Campus Goiabeiras,\\ 
29060--900 VITORIA--ES (Brasil);\\ 
2) Solid State Physics Laboratory, \\ 
Nijenborgh 4, \\ 
9747--AG GRONINGEN \\ 
(The Netherlands).} 
\maketitle 
 
\begin{abstract} 
We present a study of the nearest--neighbor (nn) and 
next--nearest--neighbor (nnn) exchange constants between magnetic Cu 
centers of the spin--Peierls material CuGeO$_3$. The dependence of these 
constants on the lattice parameters (modified e.g. by variation of 
temperature, pressure or doping) is calculated. Based on the observation 
that the bond angles are more susceptible than the bond lengths we propose 
the so--called ''{\it accordion}'' model for the description of the 
properties of CuGeO$_3$. We show that the $nn$ exchange constant in the CuO$%
_{2}$ ribbon is very sensitive to the presence and position of the side 
group Ge with respect to this ribbon. The angle between the two basic units 
the CuO$_{2}$ ribbon and the GeO$_{3}$ zig--zag chain is, besides the 
Cu-O-Cu angle in the ribbon, one of the principal lattice parameters 
determining the $nn$ exchange in the $c$ direction. The microscopic 
calculations of different exchange constants and their dependence on the 
lattice parameters are carried out using different schemes (perturbation 
theory; exact diagonalization of Cu$_{2}$O$_{2}$ clusters; band approach). 
The results compare favorable with experiment. The influence of Si doping 
is also calculated, and the reasons of why it is so efficient in suppressing 
the spin--Peierls phase are discussed. Thus the consistent microscopic 
picture of the properties of CuGeO$_{3}$ emerges. 
\end{abstract} 
 
\vfill PACS: 75.10, 75.30E, 75.30H.\newline 
(Submit to Phys. Rev. B )\newline 
Short title: spin--Peierls transition in CuGeO$_3$.  
 
\section{INTRODUCTION} 
 
\label{section: introduction} The compound CuGeO$_3$ shows a spin--Peierls 
(SP) transition at 14 K: in each Cu chain the Cu cations dimerize and the 
spins form singlets. In this paper we study the exchange constants as a 
function of variations in the lattice parameters in the uniform spin phase 
as well as the SP phase. Changes in the lattice parameters can also be 
caused by applying pressure, or doping. 
 
The plan of this paper is as follows. In the next subsections of this 
introduction we discuss the relevant structural, magnetic and magnetoelastic 
data. We finish this introduction with a general qualitative discussion of 
the theory of superexchange applied to this specific compound. 
 
In the following sections we give a detailed account of the calculation of 
the exchange constants. First, in section \ref{section: theory} we give 
details of a fourth--order perturbation theory approach. We show that the 
usually neglected side groups (in the case of CuGeO$_3$ this is Ge) have in 
this particular case a large influence on the sign and magnitude of the 
nearest--neighbor ($nn$) exchange interaction. The influence of this Ge 
side group is studied. 
 
Using perturbation theory we calculate the $nn$ (sections \ref{subsection: 
nn exchange perturbation theory} and \ref{subsection: explicit side group 
hybridization}) and next--nearest--neighbor ($nnn$) exchange (section \ref 
{section: nnn a+b+c exchange}) constants in the $c$ (chain) direction as 
well as in the $a$ and $b$ direction. The values of the exchange constants 
thus obtained are quite reasonable, but their dependence on the lattice 
parameters strongly deviates from the experimental results. Therefore, in 
section \ref{section: cluster model} we use a different method, calculating 
the states of a Cu$_{2}$O$_{2}$ plaquette exact. This method gives quite 
satisfactory results for the lattice parameter (e.g. pressure and 
temperature) dependence of these exchange constants. 
 
In order to account for multiple spin transfer paths we also performed a 
calculation using a band model for the anion states. The results reported in 
section \ref{section: band model} of the latter compare well with those 
obtained from the exact diagonalization of the Cu$_{2}$O$_{2}$ plaquette in 
section \ref{section: cluster model}. Finally, in section \ref{section: 
discussion} we give a discussion of the results; the conclusions and summary 
can be found in the final section \ref{section: conclusions and summary}. 
 
\subsection{The structure.} 
 
\label{subsection: structure} Before we discuss the magnetic properties it 
is convenient to describe relevant structural details. The structure of CuGeO%
$_{3}$ in the high temperature (HT) phase and in the SP phase has been 
determined by Braden et al \cite{Braden_et_al_(1996)}. Some results of their 
analysis can be found in table \ref{table: structure}. In fig. \ref{figure: 
CuO-chain+Ge} we have indicated the principal structural parameters 
important for our discussion of the exchange constants in the $c$ direction. 
In the HT phase all Cu and all Ge ions are equivalent. The Cu have a near 
ideal square--planar coordination by O2 and these edge--sharing squares form 
a CuO$_{2}$ ribbon in the $c$ direction. There are two O1 ions at a distance 
2.755 $\AA $ nearly perpendicular to the ribbon, which would complete a 
tetragonally distorted octahedral coordination of Cu. This distance is too 
large to be of importance for the local electronic structure of Cu. The Ge 
have a nearly ideal tetrahedral coordination of O: two O2, shared with two 
CuO$_{2}$ ribbons, and two O1's. The $\angle $ O--Ge--O angles are about 109$%
^{0}$, very near to the ideal tetrahedral angle. The interatomic distance of 
the two O2 neighbors of a Ge is: 2.82 $\AA $. These two O2 connect two Cu 
in different ribbons along the $b$--axis. This is illustrated in fig. \ref 
{figure: nn and nnn exchange b-axis} 
 
Our preferred description of this system is as follows: linear GeO$_3$ 
chains of corner sharing GeO$_4$ tetrahedra, and CuO$_2$ ribbons of edge 
sharing CuO$_4$ squares, both units in the $c$ direction. These two units 
share the O -- which we will call the {\it hinges} of the structure -- of 
the CuO$_2$ ribbon: at each side of the CuO$_2$ ribbon one GeO$_3$ chain, 
which form a {\it neutral undulating two dimensional layered} structure in 
the $b$ direction (see fig. \ref{figure: harmonica model}. 
 
The interatomic distance of O2--O2, which connect two Cu in different 
ribbons along the $a$--axis, illustrated in fig. \ref{figure: nn and nnn 
exchange a-axis}, is 3.06 $\AA $. This is much longer than twice the usually 
used ionic radius of O$^{2-}$ for 6--fold coordination, which is about 1.40 $%
\pm $ 0.05 $\AA$, but nearly equal to twice the van der Waals radius of 
oxygen ($2 \times 1.52$ $\AA$). This signifies that the bonding between two 
layers as defined in the preceding paragraph is small, and probably of the 
van der Waals type. A much smaller ionic radius of O$^{2-}$ of about 1.25 $%
\AA $, fits the interatomic distances: Ge--O2 and Cu--O2 and the O2--O2 
forming the bridge between two Cu in the same ribbon. This small ionic 
radius for O agrees with the rule from structural chemistry that a decrease 
in the coordination number causes a decrease in the ionic radius: in our 
case O2 is coordinated by only three cations. 
 
Other important structural factors for the calculation of the superexchange 
interactions are the {\it bond angles}. The most important angle is {\it %
bridge} angle $\phi =\angle $Cu--O--Cu in the ribbon plane, which is 
slightly larger than 90$^{0}$. The deviation of this 
angle from 90$^{0}$, $\beta $,  determines the sign of the $nn$ superexchange. Another important angle 
is the {\it hinge} angle $\alpha =\angle $(O$_{2}$--O$_{2}$)$_{{\rm {bridge}}%
}$--Ge, which is about 160$^{0}$. These angles change as a function of 
pressure, doping, temperature and at the SP transition. In the SP phase the 
hinge and bridge angle have two distinct values (see table \ref{table: 
structure}) due to the dimerization of the Cu chains. A very simplified 
model for this SP transition is presented in fig. \ref{figure: model SP 
transition}. It is predominantly these angular variations which determine 
the properties of CuGeO$_{3}$. 
 
The angles which determine the transfer in the $a$ and $b$ direction are 
given in fig. \ref{figure: nn and nnn exchange a-axis} and \ref{figure: nn 
and nnn exchange b-axis}: the angles $\rho_a = \angle$(O2--O2)$_{{\rm {bridge%
}}}$ --O$_a$ and $\rho_b = \angle$(O2--O2)$_{{\rm {bridge}}}$--O$_b$, which 
are about 120$^0$ and $\theta_a =\angle$(O2--O2)$_{{\rm {a}}}$--Cu and $%
\theta_b= \angle$(O2--O2)$_{{\rm {b}}}$--Cu, which are about 110$^0$. These 
angles show only very small changes as a function of temperature and through 
the SP transition. 
 
\subsection{The magnetic properties.} 
 
\label{subsection: magnetic properties} The magnetic properties -- as well 
as properties related to it -- of CuGeO$_{3}$ in the ordered SP phase can 
well be described by the Cross--Fisher \cite 
{Cross_and_Fisher_(1979),Bonner_and_Fisher_(1964)} theory, however magnetic 
properties above the SP transition in the uniform phase do not agree with 
the usual 1D--Heisenberg spin 1/2 model with $nn$ exchange $J_{nn}$ only. 
Also a consistent interpretation of these properties based on this model 
seems impossible because it leads to widely different values for this 
exchange constant. 
 
In table \ref{table: experimental exchange constants} we have collected 
experimental values for the exchange constants of CuGeO$_{3}$. The exchange 
constants in the $c$ direction are defined by the Heisenberg Hamiltonian:  
\begin{equation} 
H_{H}=J_{nn} \sum_{i}(1+\delta (T)(-1)^{i})S_{i}S_{i+1} + J_{nnn} \sum_{i} 
S_{i}S_{i+2}. 
\end{equation} 
\label{equation: Heisenberg Hamiltonian} Some authors do not specify how 
they define their exchange constants, which can easily lead to confusion. At 
present a model which includes also $nnn$ exchange $J_{nnn}$ along the Cu 
chain, gives the most consistent set of exchange parameters. Assuming $%
J_{nn}\approx 11$ to 15 meV and the frustration parameter $\gamma =J_{nnn}$ $%
/J_{nn}\approx .24$ -- .51 an agreement can be obtained between model 
calculation and experiments: high temperature (HT) susceptibility, magnetic 
specific heat, magnon dispersion. We consider the exchange constants 
determined in \cite{Fabricius_et_al_(1998)} as the most reliable set. For 
the exchange constants in the $a$ and $b$ direction we rely on the ones 
determined by Nishi et al \cite{Nishi_et_al_(1994)}. These are rather small, 
compared with the interactions in the ribbon. Note however that from recent 
analysis of Raman magnon spectra \cite{Gros_et_al_(1998)} one finds 
indications for significant interchain superexchange interactions. This 
interchain exchange is probably also responsible for the disagreement 
between the frustration parameter determined from these spectra and from 
susceptibility and other data. Also Kuroe et al \cite{Kuroe_et_al_(1994)b} 
find a somewhat larger interchain exchange interaction in the $b$ direction 
from an analysis of Raman spectra: $J_{b}=0.50$ meV. Thus, whether the Cu 
chain can safely be regarded as a 1D chain remains questionable. 
 
In the SP phase one finds two alternating values for the $nn$ exchange in 
the Cu chain. This is usual expressed by: $J_{nn}^{\pm }=J(1 \pm \delta (T))$%
, where $J$ is the average $nn$ exchange interaction in the SP phase, and $%
J_{nn}^{\pm }$ are the two alternating values. The dimensionless quantity $%
\delta (T)$ depends on the temperature. From table \ref{table: experimental 
exchange constants} we see that the estimates of this parameter at $T=0$ K 
range from 0.014 to .3, i.e. over one order of magnitude. Gros et al \cite 
{Gros_et_al_(1998)} find the best agreement with magnon spectra for an 
intermediate value. 
 
An important feature of the $J_{nn}$--$J_{nnn}$ model is that it predicts a 
spin--gap to develop in the magnetic excitation spectrum for $\gamma >\gamma 
_{c}\approx .2412$ \cite{Castilla_et_al_(1995)}. This gap opens irrespective 
of any lattice distortion. So a singlet spin liquid phase can originate 
either from the usual SP mechanism or from frustration of the 
antiferromagnetic order due to a relative strong $J_{nnn}$ exchange 
interaction. 
 
Based on the preliminary results of a study of the dependence of the 
exchange constants on the bond angles (\cite 
{Geertsma_and_Khomskii_(1996),Khomskii_et_al_(1996)}) we proposed a 
microscopic mechanism for the SP transition in CuGeO$_{3}$: in first 
approximation we attribute the main lattice changes as due to the change in 
the soft bond angles especially the angle which the Ge--O bond makes with 
the CuO$_{2}$ ribbon -- the {\it hinge} angle ($\alpha $), and the Cu--O--Cu  
{\it bridge} angle ($\phi $) in this plaquette. 
 
In fig. \ref{figure: harmonica model} we present a simple picture of this 
model as applied to the distortion of the CuGeO$_{3}$ lattice when pressure 
is applied along the $b$--axis. The two basic units of the structure -- the 
CuO$_{2}$ ribbon and the GeO$_{3}$ chain rotate around the shared O ions: 
the hinges. This hinge angle has a strong influence on the strength of the $%
nn$ exchange in CuGeO$_{3}$. We will see that its influence on the $nnn$ 
exchange is small. 
 
\subsection{The magnetoelastic properties.} 
 
\label{subsection: magnetoelastic properties} So, of interest for our model 
are the dependencies of these exchange constants on the lattice parameters. 
B\"{u}chner et al (1996) \cite{Buchner_et_al_(1996)} have recently performed 
a study of the magnetostriction and thermal expansion of CuGeO$_{3}$ single 
crystals. They obtained for the pressure dependence of the susceptibility $%
\chi$ along the three principle axes $\delta \chi_{ii}/\delta P_{i}$ at 60 
K: the $a$--axis -2.5 \%/ GPa , the $b$--axis 5 \%/ GPa, and for pressure 
along the $c$--axis: 1 \%/ GPa. These pressure dependencies are nearly 
constant from 20 K up to 60 K. Assuming that in the paramagnetic phase at 
temperatures far above the SP transition temperature, the magnetic 
susceptibility follows a Curie--Weiss type law, one derives for the 
variation of the $nn$ exchange in the ribbon with pressure $\delta \ln 
J_{nn} = 2.3\delta \ln\chi$, so: $\delta \ln J_{nn} = -5.7$, +11.5 and +2.3 
\%/GPa, in the $a$, $b$ and $c$ direction, respectively. We assume that the 
other exchange contributions are either too small or are independent of 
pressure. 
 
In order to interpret these magnetoelastic data with structural distortions 
we have to relate them to elastic data. We use the following elastic 
constants for an interpretation of these data: $c_{11} = 66$ GPa; $c_{22} = 
24$ GPa;$c_{33} = 300$ GPa. We neglect off--diagonal elastic response. 
 
First consider pressure along the b-- and c--axes as these are most 
straightforward to interpret. Assuming that most of the distortions due to 
the application of pressure are due to changes in the bond angles one finds 
for pressure along the $c$--axis, the bridge angle is the softest structural 
parameter: $\delta \phi /\delta P)_{{\rm {c\,axis}}}=-0.4^{0}$/GPa, while 
for pressure along the $b$--axis the hinge angle is the most soft structural 
parameter: $\delta \alpha /\delta P)_{{\rm {b\,axis}}}=-5^{0}$/GPa, where 
the angle $\phi =\angle $(Cu-O-Cu) is the bridge angle, and the angle $%
\alpha =\angle $(Ge-ribbon) the hinge angle. For their values see table \ref 
{table: structure}. The effect of pressure along the $b$--axis on the 
structure is illustrated with a somewhat simplified version of fig. \ref 
{figure: harmonica model} in fig. \ref{figure: simple harmonica model}. The 
compression of the lattice due to pressure along the $b$--axis changes the 
hinge angles $\alpha $ while leaving the bond lengths nearly constant. 
 
Pressure along the $b$--axis can also cause a slight increase of the bridge 
angle, but we neglect this response of the lattice; an increase in the 
bridge angle would cause a decrease of the O2--O2 bondlength, which is 
already very small and thus very unlikely. Such an increase of the bridge 
angle would make CuGeO$_3$ more antiferromagnetic, and when we ignore this 
change in $\phi$ we even somewhat underestimate the effect $P_b$ on $J_{nn}$. 
 
Another structural distortion due to pressure along the $c$--axis would be a 
buckling of the CuO$_2$ chain. The effect of such a buckling on the exchange 
constants is probably small, and antiferromagnetic. So when we ignore this 
buckling effect we underestimate the effect of $P_c$ on $J_{nn}$. 
 
The data for pressure along the $a$--axis are more difficult to translate 
into changes of a single bond angle. Pressure along this axis would cause 
probably first a decrease of the van der Waals gap between the layers. The 
next effect would be a decrease in the thickness of the these layers. This 
thickness is determined first by the Ge--O--Ge zig--zag chain. So a 
compression of the layers can be accomplished by stretching the GeO$_3$ 
chain which will force an increase of the Cu--O--Cu bridge angles. However 
like we argued in the previous paragraph this increase of the bridge angle 
is counteracted by the repulsion of the two O2 bridge ions between two $nn$ 
Cu ions in the chain. Furthermore, an increase of the tetrahedral angle at 
The Ge which connects two CuO$_2$ ribbons will flatten even more the GeO$_3$ 
chain. The latter could well lead to an increase in the distance between two 
Ge neighbors in the $a$ direction. A variation of the hinge angle by a 
rotation around the O hinges of the two principle subunits is not 
susceptible directly to the $a$--axis compression. A simple accordion effect 
increasing the $b$--axis, i.e. a rotation around the O hinges, will not help 
in compressing the system in the $a$ direction. 
 
So altogether one cannot make a straightforward estimate of the change in 
bridge angles induced by pressure along the $a$--axis. Assuming that each 
tetrahedral O--Ge--O angle contributes half of the compression one can make 
an estimate of the change in the bridge angle. From this we expect that 
pressure along the $a$--axis will have an effect opposite to pressure along 
the $c$--axis, which would be about 2 to 3 times as large as the $c$--axis 
effect. So we derive an increase of the bridge angle of about 2$^0$ per GPa 
pressure along the $a$--axis. This is about 5 times as large as for applying 
pressure along the $c$--axis, while the change in susceptibility is only 
about 2.5 times as large. From this one might conclude that probably the 
largest change in the tetrahedral angle is the one connecting two CuO$_2$ 
ribbons, i.e. the Ge--O--Ge angle in the $b$ direction, which does not 
force a change in the bridge angle, and agrees with the 
earlier argument of constant O2--O2 distance. 
 
For a detailed analysis of these data we refer to B\"{u}chner et al (1997)  
\cite{Buchner_et_al_(1997)}. These data show that the magnetic properties 
are very sensitive to the hinge angle $\alpha =\angle$ O--O--Ge and bridge angle 
$\phi = \angle$ Cu--O--Cu, like we predicted \cite{Geertsma_and_Khomskii_(1996)}. 
 
From the above we derive that the $nn$ exchange constant in the ribbon 
varies with these bond angles as follows. B\"uchner et al (1997) finds from 
the shift of the SP transition temperature under applying pressure along the  
$c$--axis a change in the $nn$ exchange constant only due to the variation 
in the bridge angle: $\delta \ln J_{nn}/\delta \phi \approx 5.8 $ \% /$^0$, 
and applying pressure along the $b$--axis a variation in the $nn$ exchange 
only due to a variation in the hinge angle: $\delta \ln J_{nn}/\delta \alpha 
\approx 2 $ \% /$^0$. 
 
This analysis of the  variations of the exchange constant with pressure 
in terms of bond angles are based on the assumption of 
rigid bond lengths, and that only one soft bond angle varies with 
pressure applied along each principle direction ($b$, $c$). 
This simple analysis is not applicable for $P_a$: in this case 
more than one bond angle is involved. 

Another way to derive 
these variations would be to correlate the angular variation with the change 
in lattice constant as a function of temperature. For the hinge angle this 
leads to a twice as large variation and for the bridge angle to nearly the 
same variation with the lattice parameters. So we suggest that the 
uncertainty in the angular variation of the $nn$ exchange constant derived 
above has an error of about 100 \% and for the hinge angle a much smaller 
uncertainty caused by variations in the bridge angle. 
 
Furthermore B\"uchner et al (1997) find the following approximate relation 
between the variation of the SP transition temperature and the average 
susceptibility ($\overline{\chi}$) at 60 K: $[\delta T_{SP}/\delta 
P_i]/[\delta \overline{\chi}/ \delta P_i] \approx 1.5\pm 0.4 $ 10$^7$ K 
g/emu for all directions $i=a$, $b$, $c$. Let us assume that the SP 
transition temperature is proportional to the exchange constant. Then we 
have for the linear variation of the $nn$ exchange constant $\delta 
J_{nn}/\delta P_i = {\rm constant} 0.5 \times \delta \chi_{ii}/\delta P_i$, 
and we find for the relative variations of the exchange constant in the $a$,  
$b$ and $c$ direction: $\delta J_{nn} = -1.1:2.5:0.6$. Above we found for 
this relative variations from the analysis of the HT susceptibility 
approximately the same ratios: $-1.1:2.2:0.45$. 
 
 
 
\subsection{Superexchange and Structure} 
 
\label{subsection: superexchange and structure} As is well--known an 
exchange between magnetic moments in insulating magnetic compounds based on 
late transition--metals is predominantly caused by the so--called 
superexchange. This superexchange is due to the overlap of the localized 
orbitals of the magnetic electrons with orbitals of intermediate nonmagnetic 
ligands. There are many processes contributing to superexchange, which 
appear under various names in various calculational schemes. Nevertheless, 
usually the sum of these partial processes give results which follow the 
Goodenough--Kanamori--Anderson (GKA) \cite{Goodenough_(1963)} rules. These 
rules are based on orbital symmetry considerations and the assumption that 
the most important covalent bonding is $dp_{\sigma }$. According to these 
rules a 180$^{0}$ superexchange (the magnetic ion--ligand--magnetic ion 
angle is 180$^{0}$) of partially filled $d$ shells with $dp_{\sigma }$ 
bonding, is antiferromagnetic, whereas a 90$^{0}$ superexchange is weakly 
ferromagnetic. This qualitative difference between 180$^{0}$ and 90$^{0}$ 
superexchange is due to the fact that, when considering $\sigma $ bonding 
only, in the former case the orbitals of the magnetic electrons are 
overlapping with the same ligand orbital, while in the latter case the 
magnetic orbitals are overlapping with different mutually orthogonal ligand 
orbitals. Neglecting direct $dd$ interactions, the molecular orbitals, based 
on these atomic $p$ and $d$ orbitals, are orthogonal, and so there is only a 
ferromagnetic superexchange contribution. 
 
Obviously, when $\pi$ bonding and/or direct $dd$ hybridization become 
important these rules do not hold anymore. Also when in the 90$^{0}$ case 
the orthogonality of the intermediate ligand orbitals is not strict the GKA 
rules have to be modified. Such is in fact the case for the $nn$ 
superexchange along the $c$--axis in CuGeO$_{3}$, which has recently been 
discussed in detail by the present authors \cite 
{Geertsma_and_Khomskii_(1996)}. 
 
There we have shown that in the case of 90$^{0}$ superexchange one should 
not only take into account the local symmetry of the cation but also the 
symmetry at the anion position. When valence orbitals of the same type 
centered at an anion involved in spin transfer between the two cations have 
different hybridization with a side group, one can have an antiferromagnetic 
interaction instead of the ferromagnetic one predicted by the GKA rules for 
90$^{0}$ superexchange. 
 
In principle similar rules hold for the $nnn$ superexchange interactions. 
For the $nnn$ superexchange in the $c$ direction the spin from one Cu can be 
transferred over two intermediate O--$p$ orbitals to the other $nnn$ Cu. In 
this case the O--$p$ orbitals have a finite transfer with both $nnn$ 
magnetic Cu--$d$ orbitals. So these $nnn$ Cu--$d$ orbitals are not 
orthogonal, and this will lead to an antiferromagnetic superexchange 
interaction; if there is a finite transfer between two magnetic orbitals, 
kinetic exchange will usually win from the ferromagnetic direct or potential 
exchange. 
 
In the case of CuGeO$_{3}$ there are two such transfer paths (fig. \ref 
{figure: nn and nnn exchange c-axis}) between $nnn$'s along the $c$--axis. 
It is clear that this leads to a relatively large antiferromagnetic 
superexchange interaction. In the $nnn$ case the ferromagnetic 
Kramers--Anderson contribution, which is due to two--site cation--$d$ anion--%
$p$ exchange and which may cause the pure 90$^{0}$ exchange to be 
ferromagnetic, is much smaller than the antiferromagnetic kinetic 
superexchange. 
 
In section \ref{section: theory} we consider the $nn$ exchange along the $c$%
--axis, using various schemes: perturbation theory, and an exact 
diagonalization of a Cu$_{2}$O$_{2}$ plaquette. The effect of the Ge side 
group attached to the bridging O pair is taken into account using two 
approaches. 
 
In the {\it first}, most simple approach the influence of Ge is taken into 
account by a shift of one of the $p$ orbitals ($p_{y}$), and a covalent 
reduction of the matrix elements containing this orbital. The hinge angle $%
\alpha$ only appears to describe the effective hybridization between the Ge--%
$sp^{3}$ and this $p_{y}$ orbitals. 
 
In a {\it second} approach to describe the influence of the side group Ge on 
the O--$p$ levels we rotate these O-p orbitals of the O--O bridge so that one 
obtains a $\sigma $ bonding between O--$p_{y}$ and the Ge--$sp^{3}$ hybrid. 
The states in the planar Ge--O--O--Ge cluster can easy be diagonalized. This 
is described in Appendix  \ref{appendix: explicit side group effect}.
 
We study these $nn$ and $nnn$ exchange interactions as a function of the 
lattice parameters. We find that perturbation theory, giving quite 
reasonable results for the exchange constants themselves, is not able to 
describe their dependence on the bond angles as found from the analysis of 
the pressure dependence of the magnetic susceptibility. We find that an 
exact diagonalization of the Cu$_{2}$O$_{2}$ plaquette is more successful in 
this respect, and results of this approach can be found in section \ref 
{section: cluster model}. The reason that the exact diagonalization of the 
cluster model gives better results compared with the fourth--order 
perturbation expansion is that in the perturbation approach the angular 
dependence is mainly caused by the hybridization, the splitting of the 
levels being constant. In the case of an exact diagonalization the changes 
in the hybridization are partly compensated by the changes in the level 
splitting. 
 
Because of the large ratio of the $nnn$ and $nn$ exchange, the question 
arises whether one should also take into account an exchange interaction 
along the $c$--axis with neighbors further away. In order to do this 
systematically we use a band model for the electronic structure of the 
nonmagnetic anion states. Such a model has been described by one of the 
present authors \cite{Geertsma_and_Haas_(1990)}. In this model we only 
consider two cations at a distance R, each with one nondegenerate 
half--filled $d$--like state, interacting via a fully--occupied valence 
band, consisting of mainly anion $p$ states. In this case one obtains a 
closed set of equations in terms of two--particle Green functions. We have 
taken into account $pp\sigma $ as well as $pp\pi $ bonding. The effect of 
the Ge side group is easily taken into account within the simple scheme, by 
shifting the energy of the $p$ levels and by introducing a covalent 
reduction parameter for the matrix elements involving these O--$p$ orbitals. 
In this approach we only calculate the contribution corresponding to the 
kinetic exchange and correlation exchange mechanism. The contributions 
stemming from the two--site $dp$ exchange and from the anion on--site $pp$ 
exchange are calculated in fourth--order perturbation theory. In section \ref 
{section: band model} we give a short account of the application of this 
model to CuGeO$_{3}$. 
 
In section \ref{section: nnn a+b+c exchange} we consider using perturbation 
theory the exchange interactions along the $a$-- and $b$--axis. An estimate 
of these superexchange constants is of importance because these determine 
whether one can safely consider CuGeO$_3$ a 1D Heisenberg spin system or one 
has also to take into account these superexchange interactions in case they 
are of comparable magnitude to the ones along the $c$--axis. The exchange 
transfer paths are illustrated in fig. \ref{figure: nn and nnn exchange 
b-axis}, and \ref{figure: nn and nnn exchange a-axis}. Note first that in 
both cases there are two exchange transfer paths for the $nn$ interactions, 
and there is a transfer path to the $nnn$ Cu in both these directions, which 
partly coincides with the one of the $nn$ transfer paths. Because the O--O 
bond length is in the $a$--axis case much longer than in the $b$--axis case, 
it is clear that the spin transfer along the $b$--axis is stronger then 
along the $a$--axis. So we expect that the kinetic superexchange 
interactions between the CuO$_2$ ribbons in the $b$ direction are larger 
than in the $a$ direction. In the $b$ direction we expect a weak 
antiferromagnetic interaction, and in the $a$ direction an even weaker 
antiferromagnetic or even weak ferromagnetic interaction. 
 
An important clue for the microscopic picture of the SP transition stems 
from a comparison of the magnetic properties of weak doping by Zn for Cu and 
Si for Ge. One finds in both cases that a small amount of impurities has an 
appreciable lowering effect on the SP transition temperature. That small 
substitutions of Si for Ge has a large effect on the SP transition is at 
first sight difficult to understand, because Ge and Si ions are only 
indirectly involved in the $nn$ superexchange interaction in the Cu chains. 
However closer examination shows that the Si ion has a number of effects on $%
nn$ superexchange along the $c$--axis. First, because Si$^{4+}$ has a 
smaller ionic radius than Ge$^{4+}$, it causes one of the Cu--O--Cu angles 
of the CuO$_{2}$ plaquette to decrease, and secondly it may increase the 
Cu--O bond length. Furthermore, this O bonded to Si is taken out of the CuO$%
_{2}$ ribbon plane. This is illustrated in fig. \ref{figure: Si substitution}. 
All these small geometrical changes add to a relative large absolute 
decrease in the kinetic exchange coming from the Cu--O--Cu transfer path to 
which Si is attached. Due to this relative local distortion of the O--hinge, 
i.e. the hinges of the accordion model, the accordion motion becomes 
frustrated. This distortion inhibits an easy rotation of the CuO$_{2}$ 
ribbon w.r.t. the GeO$_{3}$ chain. Si is attached to two CuO$_{2}$ ribbons 
and so is twice as effective as Zn on a Cu position to destroy the ordered 
SP phase. We discuss the consequences of this substitution for $nn$ 
superexchange in the $c$ direction in section \ref{section: substitutions}. 
 
 
\section{Theory} 
 
\label{section: theory} Let us first discuss the $nn$ exchange interaction 
along the $c$--axis because these seem to be of major importance for an 
understanding of the magnetic properties of CuGeO$_{3}$. The $nn$ 
superexchange involves spin exchange over two 90$^{0}$ cation--anion--cation 
transfer paths. Expressions for the various contributions to superexchange 
for this configuration have been presented before in a discussion of effects 
due to side groups -- bonded to the bridging O ligands of two magnetic 
cations -- 
on superexchange \cite{Geertsma_and_Khomskii_(1996)}. There we emphasized 
especially these side group effects in a more general context. Below we want 
to apply these ideas in a somewhat more rigorous scheme to CuGeO$_{3}$. In 
order to illustrate the method and approximation we use let us consider in 
some detail the $nn$ 90$^{0}$ superexchange along the $c$--axis in CuGeO$_{3} 
$. The Cu ions have a $d^{9}$ configuration. The unpaired electron is in a $%
d_{x^{2}-y^{2}}$ orbital as illustrated in fig. \ref{figure: old 
quantization axis}. On the oxygen we only consider the $p_{x}$ and $p_{y}$ 
orbitals in he ribbon plane. We will only consider $\sigma $--type covalent 
mixing between this Cu--$d$ orbitals and these O--$p$ orbitals. It is defined 
by $\lambda =t_{dp}/\Delta $, where $\Delta =\epsilon _{d}-\epsilon _{p}$, 
and $t_{pd}=\langle d|H_{eff}|p\rangle $ is the transfer integral. 
 
The influence of the side group Ge is taken into account in two factors. 
First, by a shift in energy ($\delta_p$) of the atomic O $p$ levels bonded with Ge, 
and secondly by introducing a covalent reduction parameter $\eta $ for the 
various interactions involving this $p$ orbital. In this model we neglect 
the influence of the $p_{z}$ orbital hybridization with the Ge $sp^{3}$%
--like hybrid. The $p_{z}$ is perpendicular to the ribbon plane and so does 
not participate in the spin transfer process. This model is equivalent to a 
model in which only the Ge-O bonding levels are taken into account in the 
spin exchange process. Because of their high energy, the antibonding Ge-O 
bonds are not participating in the spin exchange process. 
 
In \cite{Geertsma_and_Khomskii_(1996)} we did not take into account 
explicitly that Ge is outside the ribbon plane. At room temperature the 
hinge angle is about 160$^0$, i.e. about 20$^0$ out of the ribbon plane. 
Changes in this angle as a function of temperature, pressure and phase are 
-- as we will see -- the principal causes for the changes of the $nn$ 
exchange interaction along the ribbon. And thus a study of the sensitivity 
of the exchange interactions for these induced changes in the lattice 
parameters can be important for an understanding of the mechanism of the SP 
transition in CuGeO$_3$. 
 
As we discussed in the introduction, this hinge angle is very susceptible to 
pressure along the $b$--axis. In this case the CuO$_{2}$ ribbons are rigid 
units, rotating with respect to the rigid chain of GeO$_{4}$ coupled 
tetrahedra: the accordion model (see fig. \ref{figure: harmonica model}). 
The angles and bond lengths within the CuO$_{2}$ ribbon and the GeO$_{4}$ 
tetrahedra remain the same while the CuO$_{2}$ ribbons rotate around the O 
shared with the GeO$_{4}$ tetrahedra. This is illustrated in fig. \ref 
{figure: simple harmonica model} where we present a simplified version of 
this model emphasizing the changes in the hinge angle and the bridge angle. 
Inspection of fig. \ref{figure: harmonica model} makes it clear that there 
is enough room in the structure for such a rotation of these units with 
respect to each other. 
 
\subsection{The influence of side groups.} 
 
\label{subsection: nn exchange perturbation theory} In the first 
approximation one explicitly introduces a mixing of the Ge--$sp^3$ and 
O--$p$ orbitals, depending on its angle with the ribbon. We define the 
Ge$sp^3$--O$p_y$ transfer integral by $V_{\sigma} = \langle 
sp^3|H_{eff}|p_y\rangle = \cos(\alpha)\, V_{0\sigma}$, where 
$V_{0\sigma}$ 
is the value of this matrix element when Ge is in the ribbon plane. This 
model is a simple extension of the model presented by the present authors  
\cite{Geertsma_and_Khomskii_(1996)}: the hybridization and therefore the 
covalency factor $\eta$ depends on the hinge angle $\alpha$. 
 
In this model the covalency factor defined above is to lowest order  
\begin{equation} 
\eta = 1/(1+\delta_p/\delta_{\sigma}\cos^2(\alpha) )^{1/2}, 
\end{equation} 
where $\delta_p$ is the energy shift of the $p_y$ level due to the Ge--O covalency, 
and $\delta_{\sigma}$ is the energy difference between the Ge--$sp^3$ level 
and the O--$p_y$ level. This factor $\eta$ is used to correct the matrix 
elements involving the O--$p_y$ level. For example the $pp_{\sigma}$ bonding 
in the O pair bridge between two Cu in the ribbon, becomes  
\begin{equation} 
W_{\bot}=W_{0\bot}\eta^2. 
\end{equation} 
Also because of this hybridization of the $p_y$ orbitals with the Ge 
orbitals the Cu$d$--O$p$, transfer becomes less effective by a factor $\eta$. 
 
The usual way to discuss the superexchange interaction for the 90$^0$ 
cation--anion--cation configuration is to take the axis of quantization like 
in fig. \ref{figure: old quantization axis}. In this picture one can easy 
see that for 90$^0$ there is no spin transfer possible between the two 
cations via $dp_{\sigma}$ bonding. For our arguments it is convenient to 
rotate the axis of quantization as shown in fig. \ref{figure: new 
quantization axis}. In this picture spin transfer between two $nn$ Cu 
vanishes because of an interference effect: due to the phase difference 
between transfer via the various possible transfer paths the total $nn$ 
Cu--Cu spin transfer vanishes. 
 
The principal contributions to the $nn$ superexchange within this model for 
the electronic structure are a ferromagnetic contribution due to the spin 
polarization of the ligand $p$ orbitals, caused by one of the Cu spins, 
interacting with the spin on the other Cu by an two--site $pd$ exchange. 
This gives a ferromagnetic contribution:  
\begin{equation} 
J_{{\rm {nn,KA}}}=-8\lambda^2 J_{pd}, 
\end{equation} 
A second ferromagnetic contribution comes from the Hund's rule coupling on 
the O--$p$ orbitals, and is proportional to the on--site exchange interaction  
$J_H$ on the O--$p$ orbitals:  
\begin{equation} 
J_{{\rm {Hund}}}=-4\lambda^2 J_H. 
\end{equation} 
\label{equation: Hunds rule coupling} Both these contributions are possible 
in the 90$^0$ as well as in the 180$^0$ cation--anion--cation configuration. 
In the case of 90$^0$ the two-site $J_{pd}$ is smaller than in the 180$^0$ 
case, because it is in the first case effectively a $dp\pi$--type charge 
overlap, while in the second case it is a $dp\sigma$--type charge overlap. 
These two mechanisms are held responsible for the often found ferromagnetic 
superexchange in the 90$^0$ case, because as we will see in this limit the 
antiferromagnetic contributions (nearly) vanish. 
 
When the Cu--O--Cu angle deviates from 90$^{0}$ also other mechanisms start 
to contribute to the $nn$ superexchange interaction. The most important are 
the kinetic and the correlation+ring exchange mechanisms. A brief discussion 
of these mechanisms can be found in \cite{Geertsma_and_Khomskii_(1996)} and 
a detailed account in \cite{Geertsma_and_Haas_(1990)}. A simple derivation 
gives for the contribution of the kinetic exchange mechanism:  
\begin{equation} 
J_{{\rm {kin,nn,c}}}=-16\left[ \lambda _{x}^{2}\Delta _{x}-\lambda 
_{y}^{2}\Delta _{y}\right] ^{2}/U_{d}, 
\end{equation} 
where the excitation energies include the $pp$ hybridization $W_{\bot }$, 
and the shift $\delta_p$ due to the side group Ge: $\Delta _{x}=\delta _{dp}$, and $%
\Delta _{y}=\delta _{dp}+W_{\bot }+\delta_p$. The covalency parameters include the 
geometric factors: $\lambda _{x}=\lambda \sin (\phi )\sin (\phi /2)$ and $%
\lambda _{y}=\lambda \eta \sin (\phi )\cos (\phi /2)$, where $\eta $ takes 
into account that part of the $p_{y}$ orbital is hybridized into an 
antibonding Ge--O $(sp^{3}$--$p_{y})^{*}$ orbital. One can easy check that $%
J_{{\rm {kin,nn,c}}}$ vanishes in the case of equivalent $p_{x}$ and $p_{y}$ 
orbitals. 
 
For the correlation and ring exchange mechanism one finds:  
\begin{eqnarray} 
J_{{\rm {cor,nn,c}}} &= & - 16\lambda_x^4\Delta_x^2\left(\frac{1}{2\Delta_x} 
+ \frac{1}{\Delta_{xx}}\right) - 16\lambda_y^4\Delta_y^2\left(\frac{1}{ 
2\Delta_y} + \frac{1}{\Delta_{yy}}\right)  \nonumber \\ 
&& + 8\lambda_x^2\lambda_y^2(\Delta_x+\Delta_y)^2 \left(\frac{1}{ 
\Delta_x+\Delta_y} + \frac{1}{\Delta_{xy}}\right), 
\end{eqnarray} 
where the excitation energies are $\Delta_{\mu\nu} = \Delta_{\mu} + 
\Delta_{\nu} + U_{\mu\nu}$; $U_{\mu\nu}$ is the Coulomb interaction on the 
ligand $p$ orbitals. The terms with $\Delta_{\mu\nu}$ are contributions 
involving excitations from one ligand, while the terms with $\Delta_{\mu}$ 
are due to excitations from $p$ orbitals on different ligands. The latter 
are the genuine ring exchange contributions. 
 
One can show, by expanding $J_{{\rm {cor,nn,c}}}$ in $U_{\mu\nu}/(\Delta_{%
\mu} + \Delta_{\nu})$, that it includes the Hunds rule contribution $J_{{\rm  
{Hund}}}$. The remaining contribution is:  
\begin{equation} 
J_{{\rm {cor,nn,c}}}^{\prime}= -16(\lambda_x^2\Delta_x - 
\lambda_y^2\Delta_y)(\lambda_x^2-\lambda_y^2). 
\end{equation} 
This contribution also vanishes in the case of equivalent O--$p$ orbitals. 
 
Results of this approach, using the parameters of table \ref{table: 
parameters} are presented in table \ref{table: exchange constants} for 
various values of the Cu--O--Cu angle, at room temperature and at 20 K and 
also for both angles $\phi$ (see table \ref{table: structure}) in the SP 
phase and for the average angle of $\phi$ in the SP phase. Note that the 
ratio $\gamma$ of the $nnn$ and $nn$ exchange is in the region where one 
expects a significant contribution from spin frustration on the SP 
transition. 
 
An important parameter in the SP phase is the ratio: $\delta = |J_1- 
J_2|/(J_1+J_2)$. We find approximately $\delta = 0.09$, which is in the 
range of values found in literature (see introduction). 
 
The gap in the spin wave spectrum in the SP phase is approximately given by:  
$E_{SG} = 2.1J_{0}(\mu)^{2/3}$ \cite{Cross_and_Fisher_(1979)}. We find $%
E_{SG}= 1.4$ meV, which is small compared with the experimental value of 
about 2.1 meV. The deviation from the experimental value can be due to the 
extra contribution coming from the frustration of the exchange interaction 
in the CuO$_2$ ribbon. 
 
\subsection{Explicit side group hybridization} 
 
\label{subsection: explicit side group hybridization} Another and probably 
better way to include Ge is to rotate the axis of quantization of the O--$p$ 
levels with the rotation of the Ge, so that the O-$p_{y}$ orbital always 
points to the Ge--$sp^{3}$ hybrids. 
Here we will not give the details of this approach. It is 
sufficient to mention that the calculation of the exchange interactions 
follows the same calculational scheme as for the simple side group approach 
presented in the previous subsection. 
 
In this approach the covalency parameter $\eta $ does not change with the $%
\angle $Ge--(O--O)$_{bridge}$ but one now has to solve for the O-$p_{y}$ and 
O-$p_{z}$ levels explicitly, and neglecting again the Ge--like antibonding 
($sp^{3}$--$p_{y})^{*}$ states one can easily diagonalize the problem. More 
details can be found in Appendix \ref{appendix: explicit side group effect}. 
 
The results of this attempt to improve the simple side group model presented 
in the previous section are rather disappointing: the discrepancy between 
the angular dependence of the $nn$ exchange along the $c$--axis and 
experiment increases: they differ by an order of magnitude. We conclude from 
this that when one accounts only for variations of the transfer matrix 
elements and but not for the variations in the perturbed energies of the 
electronic states, one overestimates the angular dependence of the 
superexchange. Before we give details of a nonperturbational approach, which 
corrects for this deficiency, we present first results about the 
superexchange interactions which do not have such a strong angular 
dependence: the $nnn$ superexchange along the $c$--axis, and the $nn$ and $%
nnn$ superexchange in the $a$ and $b$ direction. 
 
\subsection{The next-nearest-neighbor exchange interactions.} 
 
\label{section: nnn a+b+c exchange} Let us consider the $nnn$ superexchange 
along the CuO$_{2}$ ribbon shown in fig. \ref{figure: nn and nnn exchange 
c-axis}. We take into account only the $pp_{\sigma }$ bonding: $W_{\Vert }$. 
Thus we only need to consider the $p$ orbitals directed along the $c$--axis: 
the $p_{x}$ orbitals in this figure. There is no side group effect because 
these $p$ orbitals are perpendicular to the Ge $sp^{3}$ orbitals. The 
various contributions to the $nnn$ exchange along the $c$--axis become 
 
\begin{itemize} 
\item  The {\it kinetic} exchange:  
\begin{equation} 
J_{{\rm {kin,nnn,c}}}=-4\lambda _{x}^{4}\left[ \frac{1}{\delta 
_{dp}+W_{\Vert }}-\frac{1}{\delta _{dp}-W_{\Vert }}\right] ^{2}\delta 
_{dp}^{4}/U_{d}. 
\end{equation} 
 
\item  The {\it correlation} exchange:  
\begin{eqnarray} 
J_{{\rm {cor,nnn,c}}} &=&-4\lambda _{x}^{4}\delta ^{4}\left[ \frac{1}{%
(\delta _{dp}+W_{\Vert })^{2}(2\delta _{dp}+2W_{\Vert }+U_{p_{x}}/2)}+\frac{1%
}{(\delta _{dp}-W_{\Vert })^{2}(2\delta _{dp}-2W_{\Vert }+U_{p_{x}}/2)}%
-\right.   \nonumber \\ 
&&4\left. \frac{\delta _{dp}^{2}}{(\delta _{dp}^{2}-W_{\Vert 
}^{2})^{2}(2\delta _{dp}+U_{p_{x}}/2)}\right] . 
\end{eqnarray} 
 
\item  The {\it ring} exchange contribution:  
\begin{equation} 
J_{{\rm {ring,nnn,c}}}=-2\lambda _{x}^{4}\delta _{dp}^{4}\left[ \frac{1}{%
(\delta _{dp}+W_{\Vert })^{3}}+\frac{1}{(\delta _{dp}-W_{\Vert })^{3}}-2%
\frac{\delta _{dp}}{(\delta _{dp}^{2}-W_{\Vert }^{2})^{2}}\right] . 
\end{equation} 
 
\item  The {\it ferromagnetic} Kramers--Anderson contribution:  
\begin{equation} 
J_{{\rm {ferro,nnn,c}}}=8\lambda _{x}^{2}\left[ \frac{1}{\delta 
_{dp}+W_{\Vert }}-\frac{1}{\delta _{dp}-W_{\Vert }}\right] ^{2}\delta 
_{dp}^{2}J_{pd,c}, 
\end{equation} 
where $J_{pd,c}$ is the two--site $dp$ exchange between the $p_{x}$ orbital 
and the $d_{x^{2}-y^{2}}$ orbital. We have taken $J_{pd,c}=(J_{p\sigma 
d\sigma }+J_{p\pi d\sigma })/2$, and we approximate $J_{p\sigma d\sigma 
}\approx 4J_{p\pi d\sigma }$ in our numerical evaluation of the $nnn$ 
superexchange. 
\end{itemize} 
 
The {\it total} $nnn$ exchange along the $c$--axis is given by the sum of 
these contributions:  
\begin{equation} 
J_{{\rm {tot,nnn}}} = J_{{\rm {kin,nnn,c}}} + J_{{\rm {cor,nnn,c}}} + J_{%
{\rm {ring,nnn,c}}} + J_{{\rm {ferro,nnn,c}}}. 
\end{equation} 
 
The expressions for the $nn$ superexchange interactions in the $a$ and $b$ 
direction are equivalent to those for the $nnn$ exchange in the $c$ 
direction. Also in this case there are two transfer paths involving two 
oxygens. One has only to change the definition of some of the parameters in 
the expressions for the contributions to $nnn$ exchange in the $c$ 
direction. Concerning the geometrical factors only the Cu--O--O bond angles 
and O--O bond lengths are different. In both directions ($a$ and $b$) the 
O--O bond in the transfer paths are in a plane perpendicular to the ribbon. 
In the $a$--axis case the O--O bond is nearly perpendicular to that of the 
Ge--O bond: $\tau _{a}=82^{0}$, while for the $b$--axis case this angle is 
about $\tau _{b}=35^{0} $. So for the $a$ direction we expect a weak 
influence of the side group on the superexchange interaction while in the $b$ 
direction we expect a larger influence of the side group. The reduction 
factor is: 
 
\begin{equation} 
\eta_i= 1/(1+\delta_p/\delta_{\sigma}\cos^2(\tau_i) )^{1/2}. 
\end{equation} 
The $dp$ covalency is now given by  
\begin{equation} 
\lambda_{i,\sigma}=\lambda_{0x}\eta_i \cos (\rho_i) \sin(\phi), 
\end{equation} 
where $\rho_i$ ($i=a,b$) is the angle between the O--O pair and the Cu--O 
bond in the ribbon ($\rho_a=108^0$, $\rho_b = 111^0$). The factor $\sin 
(\phi)$ corrects for the fact that the $d_{x^2-y^2}$ is not along the Cu--O 
bond. Although the two transfer paths to the $nn$ Cu sites in another ribbon 
are not equivalent concerning their torsion -- one Cu--O--O--Cu path has a 
torsion of 0$^0$ or 180$^0$, while the other has a torsion of about 110$^0$ 
-- this is of no importance for the evaluation of the superexchange when one 
only takes into account $\sigma$ bonding. 
 
A peculiarity of the CuGeO$_{3}$ structure is that in both $a$ and $b$ 
directions there is exchange to $nnn$ Cu via {\it one} transfer path which 
is equivalent with one of the transfer paths to the $nn$ Cu. So the 
contributions to the $nnn$ superexchange in the $a$ and $b$ direction are 
the same except for a numerical factor as those for the $nn$ superexchange. 
Obviously, in the latter case the ring exchange does not contribute. Thus, 
the expressions for the $nnn$ in the $a$ and $b$ direction can be easily 
written down.  
\begin{eqnarray} 
J_{{\rm {i,kin,nnn}}} &=&J_{{\rm {i,kin,nn}}}/4,  \nonumber \\ 
J_{{\rm {i,cor,nnn}}} &=&J_{{\rm {i,cor,nn}}}/2,  \nonumber \\ 
J_{{\rm {i,ferro,nnn}}} &=&J_{{\rm {i,ferro,nn}}}/2. 
\end{eqnarray} 
These two superexchange constants $nn$ and $nnn$ are of the same order of 
magnitude.  
 
\section{Exact diagonalization of the Cu$_2$O$_2$ plaquette} 
 
\label{section: cluster model} In this section we give the results of an 
exact diagonalization of the Cu$_2$O$_2$ plaquette. The set of parameters 
used in this section are the same as used in the previous sections. We also 
take into account the $pp\pi$ bonding, and take for the ratio $%
W_{\pi}/W_{\sigma} = 0.5$. 
 
The side group effect is incorporated in the following way. We neglect in 
first instance the hybridization of the $p$ orbitals on the O bridge atoms. 
Then we need to diagonalize only the Ge--O unit. The energy of the $sp^{3}$--%
$p_{\sigma }$ bonding state is:  
\begin{equation} 
\omega =0.5\left( \Delta _{\sigma }-\sqrt{\Delta _{\sigma }^{2}+4V_{0\sigma 
}^{2}}\right). 
\end{equation} 
We obtain the following expression for the covalency parameter:  
\begin{equation} 
\eta =\left[ \sin ^{2}\alpha +a_{p}^{2}\cos ^{2}\alpha \right] ^{1/2}, 
\end{equation} 
where $a_{p}$ is the amount of O--$p$ character in the $sp^{3}$--$p_{\sigma }$ 
bond, and $\alpha $ is the hinge angle. The influence of the side group is 
now taken into account as due to two sources: a geometrical and a chemical 
one. The first is reflected in the hinge angle $\alpha $, and the second by 
the hybridization of the O--$p$ orbitals with the side group (Ge). It is 
clear that when the hinge angle is 90$^{0}$ there is no side group effect, 
regardless of the strength of the hybridization. The side group effect has a 
maximum when the hinge angle is 180$^{0}$, that is in case the Ge is in the 
CuO$_{2}$ plane. 
 
Using the parameters given in the caption of table \ref{table: exact model} 
we find for the covalency parameter $\eta = 0.925$, and for the 
shift of the O--$p$ level $\delta_p = 1.65$. We have calculated the change of the $nn$ 
exchange constant as a function of temperature, accounting for the changes 
in lattice parameters, and find that it varies from 13.5 meV at 300 K to 
13.3 meV at 20 K. We also calculate the $nn$ exchange constants in the SP 
phase and find: the Cu--Cu dimer exchange is $J_{nn}^+ = 14.04$ meV and for 
the other $nn$ exchange constant we find $J_{nn}^- = 12.82$ meV, so $%
|J_{nn}^+ -J_{nn}^- |/(2J)=0.046$. This is approximately the same value 
found experimentally. Finally, we calculated the dependence of the $nn$ 
exchange as a function of the bridge and hinge angle. We find that these are 
of the right order of magnitude compared with experiment. 
 
 
\section{The band model for the superexchange interactions in the ribbon} 
 
\label{section: band model} For a further understanding and analysis of the 
magnetic properties of CuGeO$_3$ it is of importance to know whether the 
further neighbor superexchange interactions in the $c$ direction can be 
safely neglected. Furthermore we have seen that a perturbation approach is 
not able to describe the angular dependence of the $nn$ exchange 
interactions. In the previous section we gave results of an exact treatment 
of a Cu$_2$O$_2$ plaquette, and found results for the dependence of the 
exchange constants in approximate agreement with experiment. Below we 
present a model for the calculation of the exchange interaction in which the 
nonmagnetic (anion) states are represented by bands, and of the cations we 
only take into account two sites. Basically this is the two--impurity 
Anderson model applied to non-metals. We also want to study whether a 
cluster model gives the right order--of--magnitude estimate of the exchange 
constants and their dependence on the lattice parameters. A band model is 
more appropriate for a study of the $nnn$ and next $nnn$ exchange 
interactions than cluster models because it implicitly takes into account 
transfer paths up to infinite order, and a band model is only limited by the 
number of orbitals used as a basis for the description of the bands. 
 
A general model of this kind has already been proposed by Geertsma and Haas  
\cite{Geertsma_and_Haas_(1990)}. We follow this paper 
(see Appendix \ref{appendix: two impurity AM}) and give the appropriate expressions for the 
case of CuGeO$_3$ (see Appendix \ref{appendix: band model CuGeO3}). The 
basic idea of this model is to calculate the superexchange between two 
magnetic half--filled nondegenerate orbitals which are a distance R apart by 
considering only their hybridization with fully occupied valence band 
states, which usually consists mainly out of anion states. The empty 
conduction band states are neglected. The latter usually consist mainly of 
cation $s$ states. The $d$ states of the other cations are also neglected. So 
we have effectively a two--particle system. Spin transfer can only be 
mediated by excitations from the fully occupied valence states. One can now 
write down a closed set of integral equations for the two--particle (holes) 
energies of this system. The energy difference of the singlet and triplet 
ground state gives the exchange constant ($2J$). 
 
In this model we neglect the ferromagnetic two--site $dp$ contribution and 
the contribution coming from Hund's rule mechanism in the band states. These 
contributions are calculated in a perturbation scheme, similar to that in 
section \ref{section: theory}. 
 
We proceed as follows. First we calculate the ground state energy of the 
triplet states using self-consistent second--order perturbation expansion in 
the $dp$ hybridization $t_{pd}$:  
\begin{equation} 
\Omega^T = 2(\epsilon_{0d} + U_d - \Gamma(\Omega^T)). 
\end{equation} 
The expression for the energy shift function $\Gamma$ can be found in 
Appendix \ref{appendix: two impurity AM} (equation \ref{equation: Gamma}). This solution for the triplet 
state is independent of the interatomic distance between the two Cu. There 
are also fourth--order $t_{pd}$ contributions, but these are neglected. 
These can become important when the energy of the ground state in this 
self--consistent second--order approximation is very close to the band 
states. This is not the case for CuGeO$_3$. 
 
We find two contributions to the superexchange interaction. One is due to 
the kinetic spin transfer mechanism in which in the intermediate state two 
holes are on one of the cations, and the other -- the correlation and ring 
exchange mechanism, in which the two holes are in band states. The kinetic 
exchange contribution for two Cu at a distance $R$ is:  
\begin{equation} 
J_{{\rm {kin}}}({\rm {R}) = 4S_d \frac{|\gamma(\Omega^T,R)|^2}{U_d}}, 
\end{equation} 
where $\gamma(\Omega,R)$, defined by eqn \ref{equation: gamma} in 
Appendix \ref{appendix: two impurity AM} 
is the effective $dd$ transfer integral between the two Cu--$d$ orbitals. It 
is similar to the transfer integral usually defined in the Hubbard model. In 
our model the transfer integral is a function of energy. The factor $S_d$ is 
the amount of $d$ character in the two--particle ground state. This 
normalization factor is of importance in case of large effective 
hybridization between the $d$ and band states, i.e. in case when the triplet 
state $\Omega^T$ is close to the top of the valence band. 
 
The correlation and ring exchange contributions are together:  
\begin{equation} 
J_{{\rm {cor}}}({\rm {R} ) = S_d(\Theta^S(\Omega^T,R) - 
\Theta^T(\Omega^T,R))}, 
\end{equation} 
where $\Theta$ is given by eqn \ref{equation: Theta} in Appendix 
\ref{appendix: two impurity AM}. 
 
The side group effect is taken into account in the same way as in the exact 
solution for the Cu$_{2}$O$_{2}$ plaquette discussed in the previous 
section. Next we calculate the electronic structure of the O--$p$ bands. The $%
pp$ transfer matrix elements are defined with respect to their value for a 
bridge angle of 90$^{0}$. Then we can write (see fig. \ref{figure: nn and 
nnn exchange c-axis}):  
\begin{equation} 
W_{\Vert }=W_{\sigma x}=W/(1+\sin \beta )W_{\bot}=W_{\sigma y}=W/(1-\sin 
\beta ). 
\end{equation} 
The $\pi $ bonding is a factor $r=W_{\pi }/W_{\sigma }$ smaller. 
 
Because we cannot include the important ferromagnetic contributions directly 
in this model, we treat them using perturbation theory. This is permissible 
in case we calculate the angular dependence of the $nn$ exchange. because these 
two contributions depend only weakly on the variation of the bridge and 
hinge angle. For the further neighbor exchange interactions it is less 
clear whether one can safely use this perturbation scheme for the 
ferromagnetic Hund's rule and Kramers--Anderson exchange contributions. 
 
The expressions we use for the ferromagnetic exchange due to two--site $pd$ 
exchange are for the $nn$ contribution  
\begin{equation} 
J_{nn,ferro} = -8 \lambda^2 J_{pd} , 
\end{equation} 
for the $nnn$ superexchange:  
\begin{equation} 
J_{nnn,ferro} = -8\lambda^2 \frac{W_{\sigma x}^2}{\epsilon_{0d}^2}J_{pd}, 
\end{equation} 
and for the next $nnn$  
\begin{equation} 
J_{nnnn,ferro} = -8\lambda^2 \frac{W_{\sigma x}^4}{\epsilon_{0d}^4}J_{pd}. 
\end{equation} 
For the $nn$ Hunds rule contribution  
\begin{equation} 
J_{nn, Hund} = -4 \lambda^4 J_{O,H}, 
\end{equation} 
for the $nnn$  
\begin{equation} 
J_{nnn, Hund} = -4 \lambda^4 W_{0x}W_{0y}/\epsilon_{0d}^2 J_{O,H}, 
\end{equation} 
and for the next $nnn$  
\begin{equation} 
J_{nnnn, Hund} = -4 \lambda^4 W_{0x}^2W_{0y}^2/\epsilon_{0d}^4 J_{O,H}. 
\end{equation} 
For the calculation of the exchange interactions we used a similar set of 
parameters (see table \ref{table: exact model}) as used before in the 
other calculations. The results for the $nn$ and $nnn$, presented in table 
\ref{table: exact model}, are in remarkable 
agreement with those from simple perturbation theory, and agree with 
experiment; the value of the ratio of the $nnn$ and $nn$ superexchange $%
\gamma$ thus also agrees with experiment.
The $dp$ covalency with the parameters of 
this table is about 0.24 (calculated as $\sqrt{(1-S_d)/S_d}$). The shift, 
$\delta_p$, of  the O--$p $ level is about 1.65 eV, and $\eta = 0.93$.  These values are 
approximately the same as in the calculations presented in the previous 
sections. 
 
We have also calculated the $nn$ exchange and the parameter $\gamma $ as a 
function of the bridge angle ($\beta =\phi -90^{0}$) and hinge angle ($%
\alpha $). Results of this study are presented in fig. \ref{figure: band 
model bridge angle} and fig. \ref{figure: band model hinge angle}. For 
this set of input parameters the $dp$ covalency is nearly independent of 
these angles. We see that the $nn$ exchange increases by a factor two, from 
the situation when there is no side group effect ($\alpha =90^{0}$) to the 
expected maximum side group effect ($\alpha=180^{0}$). For the chosen set of 
parameters the $nn$ exchange changes sign for $\beta \approx 0$. The $nn$ 
exchange has a minimum for $\phi <90^{0}$. If our model for the lattice 
changes induced by pressure along the $b$--axis is right, then we expect 
that the hinge angle decreases on applying pressure along the $b$--axis, and 
so the $nn$ exchange decreases and the ratio $\gamma $ of $nnn$ and $nn$ 
exchange increases. 
 
We have also studied the influence of the chemical bonding factors of the 
side group. Results are presented in fig. \ref{figure: chemical bonding 
factor Vsigma} and \ref{figure: chemical bonding factor deltasigma}. The $nn$ 
exchange constant and the ratio $\gamma $ vary strongly as a function of 
these chemical bonding factors. The first derivative of the $nn$ exchange 
with respect to the bridge angle is nearly constant, while its derivative 
with respect to the hinge angle varies strongly as a function of these 
chemical bonding factors. 
 
We have also calculated the dependence of the $nn$ exchange on the $dp$ and $%
pp$ covalency, for the set of parameters of table \ref{table: exact model}.  
We find that a 1 \% change in these parameters causes a change of 
about 0.4 ($dp$) and 0.3 ($pp$) meV in the $nn$ exchange constant. 
 
We have studied the temperature dependence of the exchange constants. We 
find that the $nn$ exchange constant varies only slightly in going from 300 
to 20 K: it increases by about 0.2 meV. In the SP phase we find a difference 
of about 1.1 meV between the weak $J_{nn}^- = 13.96 $ meV, and strong one $%
J_{nn}^+ = 15.02 $ meV. This gives $|J_{nn}^+ -J_{nn}^- |/(2J) = 0.037$. 
This agrees with the ratio obtained from the exact diagonalization of the 
cluster described in the previous section. 
 
Finally, we find that the next--$nnn$ superexchange is a factor 40 to 100 
smaller than the $nnn$ superexchange interaction. Thus it is indeed safe to 
include only first and second neighbor interactions in a chain. 
 
\section{Substitution of Si by Ge.} 
 
\label{section: substitutions} There appeared some puzzling experimental 
results on the magnetic properties of low Si doped CuGeO$_3$. Compared with 
the results of low substitution of Zn for Cu, Si is surprisingly even more 
effective in suppressing the SP phase. For the case of Zn substitution this 
suppression is obvious, because Zn effectively cuts the magnetic 
interactions in the Cu chain. For Si this is less obvious. We adopt the 
following structural model to describe how Si perturbs locally the lattice. 
We assume that Si substituted for Ge is situated at the original Ge site, 
and attracts the $nn$ O to form a more or less ideal tetrahedron, with bond 
lengths like found in SiO$_2$. The part of the lattice of importance in the 
next discussion is shown in fig. \ref{figure: Si substitution}. Various 
hybridization parameters like the $dp$, the $pp$ and the $sp^3$-$p$ have to be 
corrected for the changes in the interatomic distances. Also the change of 
the geometrical factors in the form of bond angles have to be taken into 
account. We assume that all other ions remain fixed at their positions. 
 
When we substitute Si for Ge the transfer path (u) between nearest 
neighbours in the ribbon, with Si as a side group, will differ from that 
with Ge as a side group (l). So we introduce two $dp$ covalency parameters: 
one for the Si side group: $\lambda_{\mu u}$, and one for the Ge side group:  
$\lambda_{\mu l}$, where $\mu$ refers to the $p_x$ and $p_y$ orbitals on the 
O. Also the O-$p$ Si-$sp^3$ one-electron excitation energy $\delta_{\mu u}$ 
differs from that of Ge $\delta_{\mu l}$. 
 
The expressions for the contributions to the $nn$ exchange in the ribbon 
become a somewhat more involved, but can be easy written down, and are in 
Appendix \ref{appendix: Si substitution}. We approximate the $dp_{\pi}$ 
covalency by: $\lambda_{\pi} =\lambda_{\sigma}/3 $. For the two one--site $dd 
$ exchange integrals we take 1 eV. 
 
The Si--O bond length is about .13 $\AA$ smaller than that of Ge--O. Within 
the model for the deviation proposed above we find that $\beta_u \approx 3^0$%
, and that the Cu--O and (O--O)$_{bridge}$ become both about 10 \% larger: $%
t_{pdu} \approx .9 t_{pd}$, $J_{pdu} \approx .9J_{pd}$ and $W_{\bot} 
\rightarrow 0.9 W_{\bot}$. We neglect the slight deviation from planarity of 
the Cu$_2$O$_2$ plaquette. 
 
As a result we find that $J_{c,nn}\approx 0 \pm 0.5 $ meV. So Si destroys 
the $nn$ superexchange in the Cu$_2$O$_2$ plaquette. The only coupling that 
remains is the $nnn$ over the plaquette. Compared with Zn, Si will be more 
effective in destroying the $nn$ coupling, because it is attached to two 
ribbons. Another factor which may influence the doping dependence of the SP 
transition is that Zn and Si act differently on the frustration: if one 
takes into account the $nnn$ exchange interaction, a Zn--doped chain will be 
still totally antiferromagnetic (albeit with weakened AF exchange between 
Cu's across Zn), whereas Si, with the $J_{nnn}$ included will render the 
corresponding bond either completely frustrated, or even ferromagnetic. It 
is this factor which possibly explains why Si doping is not twice but even 
three times more effective in suppressing the SP phase than the Zn ones \cite 
{Grenier_et_al_(1998)} 
 
\section{Discussion.} 
 
\label{section: discussion} The fourth--order perturbation expansion usually 
used in the calculation of superexchange constants, is also successful in 
the case of the various exchange interactions in CuGeO$_3$. We find good 
agreement of our calculations with experiment. We find a value for the 
frustration parameter (see table \ref{table: exchange constants}) which is 
in the range of values found in literature (see table \ref{table: 
experimental exchange constants}). Also the values for the two alternating 
exchange constants in the SP phase is in the range of values found in the 
literature (see also table \ref{table: experimental exchange constants}. The 
exchange in the $a$ and $b$ direction are in our perturbation scheme nearly 
two orders of magnitude smaller than the $nn$ exchange in the $c$ direction. 
 
However the results for the angular dependence of the $nn$ superexchange 
using this perturbation approach differ by nearly an order of magnitude from 
the experimental data. Also further refinements in the treatment of the O--O 
bridge and the hybridization between these O and the Ge side group give no 
improvement; on the contrary, the disagreement for the angular dependence 
with experiment increases. 
 
In this perturbation scheme the transfer matrix elements depend on these 
angles. In the formulation of the superexchange these are the only 
ingredients which depend on the angle. The excitation energies do not depend 
on the transfer integrals. However, we know that in general these also may 
depend on the size of the transfer integrals. This is usually neglected, and 
the transfer integrals and excitation energy are taken from some 
tight--binding fit to an electronic bandstructure or cluster calculation, so 
that actually one takes the energy of the perturbed states: one thus 
renormalizes the excitation energies. The energy differences of these 
perturbed states have a second order dependence on the transfer integrals. 
 
When the unperturbed energy differences are of the same order as the 
transfer integrals, these energies become very sensitive to changes in the 
transfer integrals. This is actually the case in CuGeO$_3$. 
 
In order to take into account simultaneously changes in matrix elements and 
in the excitation energies, we perform an exact diagonalization of a Cu$_2$O$%
_2$ cluster, using essentially the same model as in the perturbation 
approach. We calculated the energy difference of the singlet and triplet 
state. In this approach the results for all exchange interactions and the 
dependence of the $nn$ exchange on the bond angles compare well with 
available experimental data. The shift in excitation energy due to the 
change in bond angles compensates partly the change in the transfer 
integrals. The consequence is that the $nn$ exchange in the ribbon is much 
less sensitive to changes in the lattice parameters. 
 
We obtain rather small -- nearly two orders of magnitude smaller than the $nn 
$ exchange in the $c$ direction -- values for the exchange in the $a$ and $b$ 
direction. We also find a rather small next--$nnn$ exchange in the $c$ 
direction from our solution of the band model. Thus from our calculation it 
follows that one can view this compound as a frustrated 1D spin system with $%
nn$ and $nnn$ exchange in the $c$ direction. 
 
We calculated the consequences of the substitution of Ge by Si, using the 
perturbation approach. Due to the decrease in the bridge angle $\phi$, and 
the increase in the hinge angle $\beta$ the $nn$ exchange in the ribbon 
decreases strongly. We know that the perturbation approach overestimates the 
sensitivity of exchange for these lattice changes. However even a much 
smaller change in the $nn$ exchange, for example a decrease by only 50 \%, 
would have drastic consequences, because such a value would effectively 
frustrate the exchange interaction: $nn$ and $nnn$ exchange then become 
approximately the same, which may introduce kink in the long--range order. 
 
The values for the alternating exchange constant in the SP phase are of the 
right order of magnitude; note however that the analysis of the 
experimental data give values which differ by an order of magnitude (see 
discussion in the introduction). 
 
Another consequence of this model is an elastic one. The rotation of the two 
1D sublattices CuO$_2$ and GeO$_3$ around the shared oxygen becomes 
frustrated in case of substitution of Si for Ge: the Si--O bond is shorter 
than the Ge--O bond. Rotation of these two units would then also involve a 
change in the Cu--O bondlengths: increase of the hinge angle would increase 
the Cu--O bond length, in order to accommodate the Si--O bond length. Thus we 
expect that Si doped CuGeO$_3$ shows a less strong dependence of the 
susceptibility on applying pressure along the $b$--axis. 
 
\section{Summary and Conclusions.} 
 
\label{section: conclusions and summary} In this paper we carried out 
microscopic calculations of the exchange constants in CuGeO$_3$ and studied 
their dependence on the lattice parameters. Our treatment in this paper 
allows to explain many features of this compound, which at first glance look 
rather puzzling, such as the observed sensibility for pressure applied 
perpendicular to the CuO$_2$ ribbons, and the sensitivity of the $nn$ 
exchange along the $c$--axis to Si doping 
 
We show how the account taken of the local geometry and the side groups Ge 
and Si lead to a rather detailed microscopic picture of the distortions in 
CuGeO$_3$, both above and below the SP transition. These results are largely 
specific for this particular compound, although some of the conclusions (for 
example the role of side groups in superexchange and the importance of soft 
bond bending modes) are of a more general nature. 
 
We conclude that a perturbation approach is valid for the calculation of the 
superexchange parameters for a fixed set of lattice parameters, but when one 
wants to study the dependence of these exchange parameters on the lattice 
parameters one has to take into account also the shift of the energy levels. 
Only in the limit of very ionic compounds these shifts are too small to be 
of significance. A cluster model -- in our case Cu$_2$O$_2$ -- and also a 
band model for the anion (O) states give results that compare well with 
experiment. 
 
We conclude also that the $nn$ exchange in the CuO$_2$ ribbon is most 
sensitive to changes in the hinge angle. The other exchange interactions, in 
the ribbon and between ribbons, are rather insensitive to the side group Ge. 
The sensitivity of the $nn$ exchange to the changes in the lattice is due to 
the strong interference effects which are not present in the case of the 
other exchange interactions: $nnn$ in $c$ direction, and $nn$ in the $a$ and  
$b$ direction. This sensitivity to the side groups is also the cause of the 
large influence of Si substitution on the $nn$ exchange. 
 
The presence of the two one--dimensional sublattices: the CuO$_2$ ribbon and 
the GeO$_3$ chain, which can easy rotate with respect to each other around 
the shared O ions -- the accordion model -- is the basis for the 
understanding of the properties of this inorganic spin--Peierls compound. 
 
\section*{acknowledgments} 
 
The first author thanks CNPQ for a grant -- number: 300920/97-0 -- to 
complete this work. The work of D. Khomskii was supported by the Netherlands 
Foundation for Fundamental Study of Matter (FOM) and by the European Network 
OXSEN. \newpage 
 
\appendix 
\section{ Explicit side group treatment.} 
 
\label{appendix: explicit side group effect} 
In this appendix we give the expressions for the covalency parameters 
in case we take into account explicitly the O-Ge $\sigma$ hybridization.
On the Ge we only consider a $sp^3$ hybridized orbital pointing in 
the direction of the O in the ribbon, hybridizing with the O--$p_y$ 
orbital. The O--$p_x$ orbital points along the ribbon, and therefore
the O--$p_{x}$ orbital  are not 
hybridizing with the $sp^{3}$ hybrids. We neglect again $pp_{\pi }$ 
bonding. 
For more details see the main text.
Note that we now also have to include the $p_{z}$ orbitals on the 
O, because they have a component in the CuO$_{2}$ ribbon plane. The $p_{x}$ 
are not affected, these remain in the ribbon, perpendicular to the Ge-$sp^{3} 
$ hybrid. Because all three $p$ orbitals of O can now participate in the 
spin transfer process, the number of covalency parameters increases from 2 
to 3. 
 
In this approximation the $dp_{\sigma }$ covalency parameters $\lambda _{x}$ 
, $\lambda _{y}$ and $\lambda _{z}$ can be written as:  
\begin{eqnarray} 
\lambda _{x} &=&\lambda _{0x}\sin (\phi )\sin (\phi /2),  \nonumber \\ 
\lambda _{y} &=&\lambda _{0y}\sin (\phi )\cos (\phi /2)\cos (\alpha )\eta 
_{0},  \nonumber \\ 
\lambda _{z} &=&\lambda _{0z}\sin (\phi )\cos (\phi /2)\sin (\alpha ), 
\end{eqnarray} 
where $\lambda _{0\mu }=t_{dp\sigma }/\Delta _{\mu }$ ($\mu =x$, $y$, $z$), 
and where $\Delta _{\mu }$ is the energy difference between the $d_{xy}$ 
orbital and $p_{\mu }$: $\Delta _{x}=\delta _{dp}$, $\Delta _{y}=\delta 
_{dp}+ \delta_p +W_{y}$ and $\Delta _{z}=\delta _{dp}+W_{z}$. Furthermore the $%
pp_{\sigma }$ transfer is for the bridging O pair: $W_{y}=W_{\bot }\eta 
_{0}^{2}\cos ^{2}(\alpha )$ and $W_{z}=W_{\bot }\sin ^{2}(\alpha )$. The 
covalency parameters $\lambda _{y}$ and $\lambda _{z}$ depend on the angle $%
\alpha =\angle $(Ge--(O--O)$_{{\rm {pair--bridge}}}$. When this angle is 90$%
^{0}$, $\lambda _{y}$ vanishes, because Ge is perpendicular to the Cu$_{2}$O$%
_{2}$ plaquette. In this case there is no side group effect. When $\alpha 
=180^{0}$, $\lambda _{z}$ vanishes, because Ge is in the plane of the 
plaquette. In this limit the side group effect has a maximum. 
 
In order also to take into account the change in the O--O pair bond length, 
we define $W$ as the transfer integral in case that the $\angle $ Cu--O--Cu 
is 90$^{0}$. For the $pp$ hybridization we assume a $R^{-2}$ scaling, and 
for the $dp$ we assume a $R^{-3}$ scaling So for the deviation from 90$^{0}$%
, given by $\beta $, and assuming that the Cu--O bond length does not change 
we can write for the $pp$ transfer integral of the O--O pair bridge,  
\begin{equation} 
W_{\bot }=W/(1-\sin (\beta )), 
\end{equation} 
and for later use the $pp_{\sigma }$ transfer along the ribbon  
\begin{equation} 
W_{\Vert }=W/(1+\sin (\beta )). 
\end{equation} 
 
\section{Two impurity Anderson Model for Superexchange in Non metals} 
 
\label{appendix: two impurity AM} In this appendix we give the expressions 
for the various functions appearing in equations for superexchange in CuGeO$%
_3$ in the two--impurity Anderson Model. In this model the nonmagnetic 
states are described by band states, and the magnetic states by localized $d$%
--like functions. The hybridization between these two kind of states is $t_{d%
{\bf k}}$. The only Coulomb interaction one takes into account is the 
on--site Coulomb interaction $U_d$ on the two magnetic ions. 
 
The basic equations for the electronic structure and the exchange 
interactions can be found in \cite{Geertsma_and_Haas_(1990)}. There are two 
contributions to superexchange: the kinetic superexchange and the 
correlation exchange. The latter includes the ring exchange contribution. 
The ferromagnetic contributions due to the Hund's rule coupling in the 
nonmagnetic states and the ferromagnetic due the exchange coupling between 
the band and localized states do not appear in this model. 
 
The expression for the kinetic exchange up to fourth order in the $d{\bf k}$ 
hybridization is for two $d$ orbitals 1 and 2 at a distance $R$:  
\begin{equation} 
J_{kin} = 2\left(\Gamma(\Omega^{S}) - \Gamma(\Omega^{T})\right) + 4\frac{%
|\gamma_{12}(\Omega^{S}|^2}{\Omega^{S} + 2\epsilon_{0d} + U_d - 
\Gamma(\Omega^{S})},  \label{equation general kinetic energy band model 1} 
\end{equation} 
where the triplet energy is given by the solution of:  
\begin{equation} 
\Omega^{T} = -2\left(\epsilon_{0d} + U_d - \Gamma(\Omega^{T})\right), 
\label{equation: general triplet energy} 
\end{equation} 
and the singlet energy is given by the solution of  
\begin{equation} 
\Omega^{S} = -2(\epsilon_{0d} + U_d - \Gamma(\Omega^{S})) - 2\frac{%
|\gamma_{12}(\Omega^{S})|^2}{\Omega^{S} + 2\epsilon_{0d} + U_d - 
\Gamma(\Omega^{S})}.  \label{equation: general singlet energy} 
\end{equation} 
The energy of the triplet state is in this contribution independent of the 
distance between the cations, while the energy of the singlet state depends 
on the distance between the two atoms. The second--order energy shift is:  
\begin{equation} 
\Gamma(\Omega) = \sum_{{\bf k}} \frac{|t_{d{\bf k }} |^2} {\Omega + 
\epsilon_{0d} + U_d + \epsilon_{{\bf k}}},  \label{equation: Gamma} 
\end{equation} 
and we have defined:  
\begin{equation} 
\gamma_{12}(\Omega) = \sum_{{\bf k}} \frac{t_{1{\bf k }} t_{{\bf k2}}} {%
\Omega + \epsilon_{0d} + U_d + \epsilon_{{\bf k}}},  \label{equation: gamma} 
\end{equation} 
where the sum over ${\bf k}$ also is over the bands. 
 
An expansion around the lowest triplet energy determined from equation \ref 
{equation: general triplet energy} gives for this kinetic energy 
contribution for two $d$ orbitals 1 and 2 at a distance $R$:  
\begin{equation} 
J_{kin}(R) = 4S_d \frac{|\gamma_{12}(\Omega^{T}|^2}{ U_d }, 
\label{equation: general kinetic energy band model 2} 
\end{equation} 
where $S_d$ is the total $d$ character in the particle triplet ground state. 
It is given by  
\begin{equation} 
S_d = 1/(1-2\left.\frac{\delta \Gamma(\Omega)}{\delta \Omega} 
\right|_{\Omega=\Omega^T}).  \label{equation: general renormalization factor} 
\end{equation} 
Especially in covalent systems, where the perturbed partially filled 
localized states are near the top of the fully occupied valence band, this 
factor is important to obtain good agreement between an exact solution and 
this approximate expansion. 
 
The energy of the two--particle singlet and triplet states including the 
correlation and ring exchange mechanism are up to fourth order for two $d$ 
states 1 and 2 determined by the following integral equations:  
\begin{eqnarray} 
\Omega^{S/T} &=& -2(\epsilon_{0d} + U_d - \Gamma(\Omega^{S/T})) - \sum_{{\bf %
k,l}} \left[\left( \frac{1}{\Omega^{S/T} + \epsilon_{{\bf k}} + 
\epsilon_{0d} + U_d} + \frac{1}{\Omega^{S/T} + \epsilon_{{\bf l}} + 
\epsilon_{0d} + U_d}\right)^2 \right.  \nonumber \\ 
&& \left.\frac{\left| t_{1{\bf l}}\right|^2 \left| t_{2{\bf k}} \right|^2 
\pm t_{2{\bf l}} t_{{\bf l 1}} t_{1{\bf k}} t_{{\bf k 2}}} {\Omega^{S/T} + 
\epsilon_{{\bf k}} + \epsilon_{{\bf l}}}\right],  \label{equation: Theta} 
\end{eqnarray} 
where the +/- apply to the singlet/triplet state. Let us denote the sum over  
${\bf k,l}$ in the rhs of this expression by $\Theta^{S/T}$. We assume that 
the exchange constant is much smaller than any of the excitation energies. 
We may expand the energy of the singlet state around the triplet state 
energy given by equation \ref{equation: general triplet energy}, and we 
obtain for the correlation contribution to the superexchange interaction:  
\begin{equation} 
J_{cor}(R)= S_d ( \Theta_{12}^S - \Theta_{12}^T), 
\label{equation: correlation exchange bandmodel} 
\end{equation} 
which can be further approximated by  
\begin{equation} 
J_{cor}(R) = 2S_d \sum_{{\bf k,l}} \left[\left( \frac{1}{\Omega^{T} + 
\epsilon_{{\bf k}} + \epsilon_{0d} + U_d} + \frac{1}{\Omega^{T} + \epsilon_{%
{\bf l}} + \epsilon_{0d} + U_d}\right)^2 \frac{t_{2{\bf l}} t_{{\bf l1}} t_{1%
{\bf k}} t_{{\bf k2}}} {\Omega^{T} + \epsilon_{{\bf k}} + \epsilon_{{\bf l}}}%
\right].  \label{equation: approximate correlation band model} 
\end{equation} 
 
Using one single orbital for the sum over ${\bf k}$ and ${\bf l}$ one can 
easily check that this leads to the usual expression for the correlation 
exchange contribution for a three center cation--anion--cation cluster 
model. Also one can check easily that it includes the ring exchange 
contribution. 
 
\section{Application to CuGeO$_3$} 
 
\label{appendix: band model CuGeO3} In the case of CuGeO$_3$ we only 
consider the CuO$_2$ ribbons. 
 
The band states are described by $\epsilon _{\mu {\bf k}}$. where $\mu $ is 
a band index. The unperturbed $d$ states $d_{1}$ and $d_{2}$ have an energy $%
\epsilon _{0d}$. We only consider the O--$p$ orbitals $p_{x}$, $p_{y}$ in 
the CuO$_{2}$ ribbon plane. We take into account the $pp\sigma $ as well as 
the $pp\pi $ hybridization. These split up into four bands, of which only 
two are interacting with the $d_{xy}$ orbitals. The linear combinations are:  
\begin{eqnarray} 
p_{x^{-}}=(p_{xu}-p_{xl})/\sqrt{2};  &&p_{y^{+}}=(p_{yu}+p_{yl})/\sqrt{2}, 
\end{eqnarray} 
where the subscript $u$ and $l$ refer to the upper and lower O chain. The 
energies of these two linear combinations are:  
\begin{eqnarray} 
\epsilon ^{-}=\epsilon _{px}=\epsilon _{0px}+W_{\pi x}; &&\epsilon ^{+}=\epsilon 
_{py}=\epsilon _{0py}-W_{\sigma y}. 
\end{eqnarray} 
The $W_{\sigma y}$ also contains a covalent factor $\eta $ which takes into 
account the hybridization of the $p_{y}$ orbital with the side group 
orbitals in our case the $sp^{3}$ hybrids of Ge. Also the energy of the $%
p_{y}$ orbital is shifted by an amount $\delta_p$ due to this Ge--O hybridization. 
So $\epsilon _{0py}=\epsilon _{0px}+ \delta_p$. For the $pp$ hybridization we assume 
a $R^{-2}$ scaling, and for the $dp$ we assume a $R^{-3}$ scaling. Both $%
W_{\pi x}$ and $W_{\sigma y}$ depend on the bridge angle $\phi $. Expressed 
in terms of their value for the case $\phi =180^{0}$, we can write for these 
two parameters:  
\begin{eqnarray} 
W_{\sigma x}=W_{0}/(1-\sin \beta ); &&W_{\sigma y}=W_{0}/(1+\sin \beta ). 
\end{eqnarray} 
 
For convenience we define the following quantities:  
\begin{eqnarray} 
g_{x}=[\sin (\phi )\cos (\phi /2)]^{2};& & g_{y}=[\eta \sin (\phi )\sin (\phi 
/2)]^{2}. 
\end{eqnarray} 
These expression contain the geometrical factors: the bridge angle $\phi $, 
and the influence of the side group is represented by the covalency factor $%
\eta $. The energy shift functions for the two bands read:  
\begin{eqnarray} 
\Gamma _{x^{-}}(\Omega ) &=&\frac{8t_{pd}^{2}}{2\pi }g_{x}\int_{0}^{\pi }%
\frac{1+\cos k}{\Omega +\Delta _{x}(k)}dk, \\ 
\Gamma _{y^{+}}(\Omega ) &=&\frac{8t_{pd}^{2}}{2\pi }g_{y}\int_{0}^{\pi }%
\frac{1-\cos k}{\Omega +\Delta _{y}(k)}dk, 
\end{eqnarray} 
and the total shift is  
\begin{equation} 
\Gamma (\Omega )=\Gamma _{x^{-}}(\Omega )+\Gamma _{y^{+}}(\Omega ), 
\label{equation: Gamma CuGeO3} 
\end{equation} 
where we have defined:  
\begin{eqnarray} 
\Delta _{x}(k)=\epsilon _{x}+W_{\sigma x}\cos k; &&\Delta _{y}(k)=\epsilon 
_{y}-W_{\pi y}\cos k ,  \nonumber \\ 
\epsilon _{x}=\epsilon _{0d}+U_{d}+\epsilon _{0px}+W_{\pi x}; &&\epsilon 
_{y}=\epsilon _{0d}+U_{d}+\epsilon _{0py}+W_{\sigma y} . 
\end{eqnarray} 
The kinetic transfer for two $d_{xy}$ orbitals at a distance $R=cn$, where $n 
$ is an integer, and c the $nn$ Cu--Cu distance in the ribbon, is determined 
by the function:  
\begin{eqnarray} 
\gamma _{x^{-}}(\Omega ,n) &=&\frac{8t_{pd}^{2}}{2\pi }g_{x}\int_{0}^{\pi }%
\frac{(1+\cos k)\cos kn}{\Omega +\Delta _{x}(k)}dk , \nonumber \\ 
\gamma _{y^{+}}(\Omega ,n) &=&\frac{8t_{pd}^{2}}{2\pi }g_{y}\int_{0}^{\pi }%
\frac{(1-\cos k)\cos kn}{\Omega +\Delta _{y}(k)}dk ,
\end{eqnarray} 
and the total kinetic transfer is  
\begin{equation} 
\gamma (\Omega ,n)=\gamma _{x^{-},n}(\Omega )+\gamma _{y^{+},n}(\Omega ) .
\label{equation: kinetic transfer function CuGeO3} 
\end{equation} 
The total kinetic energy shift is  
\begin{equation} 
\Theta _{kin}(\Omega ,k)=4\frac{|\gamma (\Omega ,n)|^{2}}{\Omega +2\epsilon 
_{0d}+U_{d}-\Gamma (\Omega )} .
\label{equation: kinetic energy shift function CuGeO3} 
\end{equation} 
The correlation energy shifts are for the singlet and triplet state:  
\begin{eqnarray} 
\Theta _{xx}^{S/T}(\Omega ,n) &=&\int_{-\pi }^{\pi }dk\int_{0}^{\pi }dl\frac{%
(1+\cos k)(1+\cos l)(1\pm \cos (k-l)n)}{\Omega +\Delta _{xx}(k,l)}\left(  
\frac{1}{\Omega +\Delta _{x}(k)}+\frac{1}{\Omega +\Delta _{x}(l)}\right) ^{2} ,
\nonumber \\ 
\Theta _{y}^{S/T}(\Omega ,n) &=&\int_{-\pi }^{\pi }dk\int_{0}^{\pi }dl\frac{%
(1-\cos k)(1-\cos l)(1\pm \cos (k-l)n)}{\Omega +\Delta _{yy}(k,l)}\left(  
\frac{1}{\Omega +\Delta _{y}(k)}+\frac{1}{\Omega +\Delta _{y}(l)}\right) ^{2} , 
\nonumber \\ 
\Theta _{xy}^{S/T}(\Omega ,n) &=&\int_{-\pi }^{\pi }dk\int_{0}^{\pi }dl\frac{%
(1+\cos k)(1-\cos l)(1\pm \cos (k-l)n)}{\Omega +\Delta _{xy}(k,l)}\left(  
\frac{1}{\Omega +\Delta _{x}(k)}+\frac{1}{\Omega +\Delta _{y}(l)}\right) ^{2} . 
\nonumber \\ 
&&  \label{equation: correlation energy shift function CuGeO3} 
\end{eqnarray} 
The $+$, $-$ sign applies to the singlet (S) and triplet (T) state 
respectively. The total correlation energy shift is for the S/T two particle 
state:  
\begin{equation} 
\Theta ^{S/T}(\Omega ,n)=2\frac{t_{pd}^{4}}{\pi ^{2}}\left[ g_{x}^{2}\Theta 
_{xx}^{S/T}(\Omega ,n)+g_{y}^{2}\Theta _{yy}^{S/T}(\Omega 
,n)+2g_{x}g_{y}\Theta _{xy}^{S/T}(\Omega ,n)\right] . 
\label{equation: total correlation energy shift function CuGeO3} 
\end{equation} 
From this quantity for $n=1$, 2 and 3 we calculate the $nn$, $nnn$ and next $%
nnn$ superexchange interactions.  
 
\section{ The equations for the $nn$ exchange with Si substitution.} 
 
\label{appendix: Si substitution} 
In this appendix we give the expressions for the $nn$ exchange in the $c$ direction 
in case Si is sunstituted for one of the Ge side groups. 

The {\it ferromagnetic} $dp$ contribution 
is given by:  
\begin{eqnarray} 
J_{{\rm {ferro}}} &=& 4[\lambda_{0xl}^2\cos^6(\beta_l) + 
\lambda_{0xu}^2\cos^6(\beta_u)]J_{d\sigma p\pi}  \nonumber \\ 
&& + 4[\lambda_{0xl}^2\sin^2(\beta_l)\cos^4(\beta_l) + 
\lambda_{0xu}^2\sin^2(\beta_u)\cos^4(\beta_u)]J_{d\sigma p\sigma}, 
\end{eqnarray} 
where we have neglected the $J_{d\pi p\pi}$ and the $J_{d\pi p\sigma}$ 
proportional to $\sin^4(\beta)$. 
 
The {\it kinetic} exchange contribution is  
\begin{equation} 
J_{{\rm {kin}}}=-4/U_d \left[\lambda_{xl}^2\Delta_{xl} + \lambda_{xu}^2 
\Delta_{xu} - \lambda_{yl}^2\Delta_{yl} - \lambda_{yu}^2\Delta_{yu}\right]^2. 
\end{equation} 
 
There are two contributions to the correlation exchange. The first one 
involves a double excitation on one O and one $p$ level, and gives  
\begin{equation} 
J_{{\rm {cor,1}}}= -8\lambda_{xu}^4\Delta_{xu}^2/D_{xxu} 
-8\lambda_{xl}^4\Delta_{xl}^2/D_{xxl} -8\lambda_{yu}^4\Delta_{yu}^2/D_{yyu} 
-8\lambda_{yl}^4\Delta_{yl}^2/D_{yyl}. 
\end{equation} 
 
The second contribution is due to the double excitation from two $p$ levels 
on the same O:  
\begin{equation} 
J_{{\rm {cor,2}}}= 4\lambda_{xu}^2\lambda_{yu}^2(\Delta_{xu} + 
\Delta_{yu})^2/D_{xyu} + 4\lambda_{xl}^2\lambda_{yl}^2(\Delta_{xl} + 
\Delta_{yl})^2/D_{xyl} .
\end{equation} 
 
Finally we have the {\it ring} exchange contribution, which is given by the 
double excitation from $p$ orbitals on different O of the O pair bridge. One 
obtains:  
\begin{equation} 
J_{{\rm {ring}}} = -4\lambda_{xu}^2\lambda_{xl}^2(\Delta_{xu} + \Delta_{xl}) 
-4\lambda_{yu}^2\lambda_{yl}^2(\Delta_{yu} + \Delta_{yl}) 
-4\lambda_{xu}^2\lambda_{yl}^2(\Delta_{xu} + \Delta_{yl}) 
-4\lambda_{yu}^2\lambda_{xl}^2(\Delta_{yu} + \Delta_{xl}) .
\end{equation} 
Note that these correlation contributions already contain the contribution 
due to Hunds rule. 
 
Next to these contributions there appear two new ferromagnetic contributions 
due to a kinetic transfer mechanism. The contributions vanish in the case 
that the transfer paths along the two bridge O are the same. In the case of 
a single Si substitution the upper (via the O bonded to Si) and lower 
(bonded to Ge) transfer paths are inequivalent, The two contributions which 
appear are due to Hunds rule coupling on the Cu: one due to $d_{\pi}$--$%
d_{\sigma}$ and the other due do $d_{\sigma_1}$--$d_{\sigma_1}$ on--site 
exchange. 
 
We find for these contributions:  
\begin{equation} 
J_{nn\pi\sigma} = 4\left[ (\lambda_{xu}\lambda_{\pi xu} + 
\lambda_{yu}\lambda_{\pi yu} )\Delta_u - (\lambda_{xl}\lambda_{\pi xl} + 
\lambda_{yl}\lambda_{\pi yl} )\Delta_l \right]^2 \left(\frac{J_{\pi \sigma}}{%
U_d}\right)^2 ,
\end{equation} 
and  
\begin{equation} 
J_{nn z^2,x^2-y^2} =4\left[(\lambda_{xu}\lambda_{z^2 xu} + 
\lambda_{yu}\lambda_{z^2 yu} )\Delta_u - (\lambda_{xl}\lambda_{z^2 xl} + 
\lambda_{yl}\lambda_{z^2 yl} )\Delta_l \right]^2 \left(\frac{J_{z^2,x^2-y^2}%
}{U_d}\right)^2 .
\end{equation}

\pagebreak

\begin{table}[tbp] 
\caption[Structure details]{Structure details relevant for the calculation 
of the various superexchange interactions. Data are from \protect\cite 
{Braden_et_al_(1996)}. } 
\label{table: structure} 
\begin{tabular}{|r|c|c|c|} 
\hline 
Bond/angle & 300 K & 20K & 4.2 K \\  
\hline 
Cu--O2 & 1.9326 & 1.9327 & 1.9351/1.9322 \\  
(O2--O2)$_{{\rm {\ perp. chain}}}$ & 2.5089 & 2.504 & 2.4984/2.5159 \\  
(Ge--O1)$_{{\rm {chain}}}$ & 1.7730 & 1.7761 & 1.742 \\  
Cu--O1$_{{\rm {apical}}}$ & 2.7549 & 2.7295 & 2.7300 \\  
(O2--O2)$_{{\rm {a--axis}}}$ & 3.062 & 3.028 &  \\  
(O2--O2)$_{{\rm {b--axis}}}$ & 2.8249 & 2.825 &  \\  
(O2--O2)$_{{\rm {par.chain}}}$ & 2.9404 & 2.9445 &  \\ \hline 
$\alpha = \angle$Ge--(O--O)$_{{\rm {c--bridge}}}$ & 159.52 & 158.85 &  
159.86/158.10 \\  
$\phi = \angle$(Cu--O--Cu)$_{{\rm {ribbon}}}$ & 99.06 & 99.24 & 99.56/98.76 
\\  
$\tau_a = \angle$Ge--(O--O)$_{{\rm {a--bridge}}}$ & 82.00 & 81.55 &  \\  
$\tau_b = \angle$Ge--(O--O)$_{{\rm {b--bridge}}}$ & 35.38 & 35.49 &  \\  
$\rho_a = \angle$(O--O)$_{{\rm {c--bridge}}}$--O$_{{\rm {a path}}}$ & 118.48 
& 119.62 &  \\  
$\rho_b = \angle$(O--O)$_{{\rm {c--bridge}}}$--O$_{{\rm {b path}}}$ & 124.14 
& 123.35 &  \\  
$\theta_a = \angle$(O--O)$_{{\rm {a--bridge}}}$--Cu & 108.03 & 108.67 &  \\  
$\theta_b = \angle$(O--O)$_{{\rm {b--bridge}}}$--Cu & 111.36 & 110.81 &  \\  
$\angle_{{\rm {torsion}}}$(Cu--(O--O)$_a$--Cu) & 72.95 (1$\times$) &  &  \\  
& 180.0 (1$\times$) &  &  \\  
$\angle_{{\rm {torsion}}}$(Cu--(O--O)$_b$--Cu) & 109.22 (1$\times$) &  &  \\  
& 0.00 (1$\times$) &  &  \\ \hline 
\end{tabular} 
\end{table} 
\pagebreak   
\begin{table}[tbp] 
\caption[table exchange constants]{Some experimental values for the exchange 
constants in CuGeO$_{3}$, all in meV. The parameters are defined by the 
equation \ref{equation: Heisenberg Hamiltonian}. } 
\label{table: experimental exchange constants} 
\begin{tabular}{|cccc|c|c|l|} 
\hline 
J$_{nn}$ & J$_{nnn}$ & J$_{a}$ & J$_{b}$ & $\gamma$ & $\delta(0)$ & method 
reference \\ \hline\hline 
-10.4 & -- & 0.044 & -0.35 & -- & -- & Magn. Suscept.; \cite 
{Kuroe_et_al_(1994)} \\  
-7.6 & -- & -- & -- & -- & 0.17 & Magn. Suscept.;\cite{Hase_et_al_(1993)} \\  
-10.4 & -- & 0.1 & -1.0 & -- & 0.12 & Inel.Neutron Scat.; \cite 
{Nishi_et_al_(1994)} \\  
-15.5 & -- & -- & -- & -- & -- & Magn. sat.; \cite{Nojiri_et_al_(1997)} \\  
-11.9 & -- & -- & -- & -- & -- & Low field magn.; \cite{Hase_et_al_(1996)} 
\\  
-5.8 & -4.9 & 0.00$\pm$0.03 & -0.315 & 0.86 & -- & Magnon disp.; \cite 
{Cowley_et_al_(1996)} \\  
-13.0 & -3.1 & -- & -- & 0.24 & 0.03 & HT Magn. Suscept.+ LT Magnon 
dispersion; \cite{Castilla_et_al_(1995)} \\  
-13.8 & -4.9 & -- & -- & 0.36 & 0.014 & HT Magn. Suscept.; \cite 
{Riera_and_Dobry_(1995)},\cite{Riera_and_Koval_(1996)} \\  
-12.8 & -6.6 & -- & -- & 0.51 & -- & Inel. Neutron Scat.; \cite 
{Arai_et_al_(1996)} \\  
-15.5 & -2.8 & -- & -- & 0.18 & -- & Raman; \cite 
{van_Loosdrecht_et_al_(1997)} \\  
-11.0 & -- & -- & -- & -- & -- & AF Res.; \cite{Hase_et_al_(1996)}, Cu(Zn)GeO%
$_3$: $J = J_{nn}+J_b$ \\  
-13.8 & -4.9 & -- & -- & 0.36 & -- & HT Magn. Suscept. + magn.--stric.; \cite 
{Fabricius_et_al_(1998)} \\  
-10.7 & -3.7 & -- & -- & 0.35 & -- & HT Magn. Spec. Heat from Raman Scat.;  
\cite{Kuroe_et_al_(1997)} \\  
-9.02 & -- & 0.050 & -0.50 & -- & -- & Raman spectra: \cite 
{Kuroe_et_al_(1994)b}. \\ \hline 
\end{tabular} 
\end{table} 
\pagebreak

\begin{table}[tbp] 
\caption[The magnetoelastic constants]{The variation of the magnetic 
susceptibility with pressure from \protect\cite{Buchner_et_al_(1997)}, the 
lattice constants with pressure from: \protect\cite{Buchner_private}. In the 
final column we find the variation of the principle bond angle.} 
\label{table: magnetoelastic constants} 
\begin{tabular}{|c|cccc|} 
\hline 
Direction i & $ delta \ln \chi_i/\delta P_i$ & $c_{ii}$ & $\delta \ln 
i/\delta P_i$ & $\delta J_{nn}/\delta P_i$ \\  
& [\%/GPa] & [GPa] & [\% /GPa] & [meV/GPa] \\ \hline\hline 
a & -2.5 & 66 & -1.5 & -0.60 \\  
b & +5 & 24 & -4.2 & +1.20 \\  
c & +1 & 300 & -0.3 & +0.24 \\ \hline 
\end{tabular} 
\end{table} 
\pagebreak 
\begin{table}[tbp] 
\caption[Parameter values]{Parameter values used in the calculation of the 
superexchange interaction. Energies are in eV.} 
\label{table: parameters} 
\begin{tabular}{|c|c||c|c||c|c||c|c|} 
\hline 
$t_{dp\sigma}$ & 1 & $U_d$ & 7 & $U_{pxx}$ & 5 & $J_{pd\pi} $ & 0.025 \\  
$W_{pp\sigma}$ & 1 & $J_{Hund,O}$ & 0.4 & $U_{pyy}$ & 5 & $J_{dz^2,dx^2-y^2}$ 
& 1 \\  
$\delta_{dp}$ & 4.0 & $J_{d\pi d\sigma}$ & 1 & $U_{pxy}$ & 4.2 & $\delta_p$ & 0.4 
\\ \hline 
\end{tabular} 
\end{table} 
\pagebreak 
\begin{table}[tbp] 
\caption[The exchange constants]{The exchange constants as calculated in the 
various perturbation approximations described in the text.
We give only the parameters which differ from the ones given in table 
\ref{table: parameters}.
\newline 
A: Perturbation results for the simple Ge--O hybridization model. The 
geometrical dependence of the two--site pd exchange is not explicit taken 
into account. Parameters: 
$t_{pd}=0.9$;
$\delta_{\sigma} = 4$; 
$J_{apd}= 0.017$; 
$J_{bpd}= 0.018$; 
$J_{cpd}= 0.029$.  
\newline 
B: Perturbation results for the simple Ge--O hybrization model for the Cu$_2$%
O$_2$ cluster. Parameters: 
$t_{pd}=0.9$; 
$\delta_{\sigma} = 4$; 
$J_{p\sigma d\sigma} = 3J_{p\pi d\sigma}$; 
$J_{p\pi d\pi} =0$; 
$J_{p\sigma d\pi} = 0.5J_{p\pi d\sigma}$. \newline 
C: Perturbation result for the Cu$_2$O$_2$Ge$_2$ cluster, the $p_y$--$p_z$ 
mixing is neglected. Parameters: 
$\delta_{\sigma} = 4$; 
$J_{cpd} = 0.026$ (0.053); 
$J_{apd} = 0.031$;  
\newline 
$J_{bpd} = 0.032$; 
$J_{xpd} = 0.053$; 
$J_{ypd} = 0.045$.\newline } 
\label{table: exchange constants} 
\begin{tabular}{|c|c|c|c|c|} 
\hline 
Model & A & B & C & experiment\\ \hline\hline 
$J_{c,nn}$ 300 K & 9.8 & 102 & 7.8&13.8 \cite{Fabricius_et_al_(1998)}  \\  
\hfill 20 K & 10.1 & 10.6 & 8.0          & --\\  
\hfill 4 K (1) & 10.8 & 11.2 & 9.2      & --\\  
\hfill (2) & 9.0 & 9.6 & 6.4                 &   --\\  
$J_{c,nnn}$ & 3.2 & 3.2 & 5.2 (4.20)& 4.9 \cite{Fabricius_et_al_(1998)}     \\  
$J_{a,nn}$ & 0.02 & -- & -0.03         &   -0.1\cite{Nishi_et_al_(1994)}  \\  
$J_{a,nnn}$ & -0.02 & -- & -0.07      & --\\  
$J_{b,nn}$ & 0.06 & -- & 0.02          & 1.0 \cite{Nishi_et_al_(1994)} \\  
$J_{b,nnn}$ & -0.02 & -- & -0.08      & --\\ \hline 
$\delta \ln J_{c,nn}/\delta\beta$ & 20 \% & 18 \% & 30--45 \%  &  5.8  \%\\  
$\delta \ln J_{c,nn}/\delta\alpha$ & 0.5\% & 0.5 \% & 4.8  \%    &  2.0  \% \\  
$\gamma$ (20 K ) & 0.32 & 0.3 & 0.66(0.52)                  &  0.36\cite{Fabricius_et_al_(1998)}        \\  \hline 
\end{tabular} 
\end{table} 
\pagebreak  
\begin{table}[tbp] 
\caption[table: exact model]{The exchange constants for the 
exact solution of: \newline 
The Cu$_2$O$_2$ plaquette. Parameters used: $t_{pd} = 1.0$%
, $W = .9$; $\epsilon_{0d} = -4.0$; $U_d = 7$, $V_{0\sigma} = 3.7$; $%
\Delta_{\sigma} = 6.65$; $J_{pd} = 0.03$; $\epsilon_{0px} = -13.4$; $U_x = 
U_y = 5.0$; $U_{xy} = 3.8$. \\ 
Band model: 
Parameters used are the same as for the plaquette except:  
$\epsilon_{0d} = -3.7$; $J_{pd} = 0.04$; $J_{OH} = 0.6$. 
All energies are in eV.  Exchange constants in meV.
The structural parameters we used are those at 20 K. } 
\label{table: exact model} 
\begin{tabular}{|l|cccccc|} 
\hline 
Model& $\alpha$ & $\beta$ & $J_{nn}$        &$\gamma$ & $\delta\ln J_{nn}/\delta \beta$ & $\delta\ln 
J_{nn}/\delta \alpha $ \\ \hline 
Cu$_2$O$_2$ plaquette &158.85 & 9.24 & 13.5 & --   & 9.7 \% & 0.56 \% \\  
\hline  
band model                     &158.85 & 9.24 & 14    & .31  &  10 \% & 1 \% \\  
\hline 
experiment & --     &    --  &   13.8 \cite{Fabricius_et_al_(1998)}    & 0.36 \cite{Fabricius_et_al_(1998)}   & 5.8  \%  &   2 \%  \\   \hline
\end{tabular} 
\end{table} 
 
\pagebreak  
 
\clearpage 
\newpage  
\begin{figure}[tbp] 
\caption[fig 1]{The CuO$_2$ ribbon, with the Ge side groups, and the 
tetragonal distortion of the Cu coordination by O: the four O2 form a 
rectangular coordination of Cu, while the two O1 are shifted along the 
"tetragonal" axis rather far away from Cu. The bridge angle $\phi$ and hinge 
angle $\alpha$ are indicated.} 
\label{figure: CuO-chain+Ge} 
\end{figure} 
\begin{figure}[tbp] 
\caption[fig 2]{The $nn$ and $nnn$ transfer paths for superexchange in the $b 
$ direction. The $nn$ exchange in the $b$ direction is between Cu--0 and 
Cu--1, and the $nnn$ exchange is between Cu--0 and Cu--1'. The bond lengths 
and angles used in the calculation are indicated. For details see the text.} 
\label{figure: nn and nnn exchange b-axis} 
\end{figure} 
\begin{figure}[tbp] 
\caption[fig 3]{The $nn$ and $nnn$ transfer paths for superexchange in the $a 
$ direction. The $nn$ exchange is between Cu--0 and Cu--1 is along two 
Cu--O--O--Cu transfer paths, while the $nnn$ to Cu--1' is along one of these 
transfer path. The various bond lengths and angles used in the text are 
indicated.} 
\label{figure: nn and nnn exchange a-axis} 
\end{figure} 
\begin{figure}[tbp] 
\caption[fig 4]{The SP transition. We emphasize the SP distortion as 
predominantly one where the angles are changing. In the SP phase the 
dimerization of the Cu chain is predominantly due to the occurrence of two 
sets of angles ($\phi_1$, $\alpha_1$) and ($\phi_2$, $\alpha_2$), the bridge 
and hinge angle respectively.} 
\label{figure: model SP transition} 
\end{figure} 
\begin{figure}[tbp] 
\caption[fig 5]{ The accordion effect due to pressure along the $b$--axis, 
viewed along the $c$--axis. The CuO$_2$ ribbon/chains as well as the GeO$_3$ 
chains are rotating with respect to each other as rigid units, around the 
shared O2 ions: the hinges. Thereby the hinge angle $\alpha$ decreases. In 
the figure we keep the GeO$_3$ chain in unit A fixed. Arrows indicate the 
movement of the neighbor units: CuO$_2$ ribbon, and the GeO$_3$ chains 
w.r.t. to this fixed unit. The direction and length of these arrows is only 
a general indication of the type of distortion.} 
\label{figure: harmonica model} 
\end{figure} 
\begin{figure}[tbp] 
\caption[fig 6]{A simplified representation of the distortion due to 
pressure along the $b$--axis. The hinges are at the O2 atoms. We emphasize 
the central role of the hinge angle in the distortion of the lattice: both 
Ge--O bonds move clock--wise, leaving the bond lengths constant; thereby the 
hinge angle $\alpha$ decreases. } 
\label{figure: simple harmonica model} 
\end{figure} 
\begin{figure}[tbp] 
\caption[fig 7]{Exchange along the $c$--axis. We have indicated the various 
transfer matrix elements used in the text: $t_{pd}$ between Cu-$d$ and O-$p$%
, $W_{\Vert}$ and $W_{\bot}$ -- the $pp$ transfer between two $nn$ O along 
and perpendicular to the Cu chain, and $V_{0\sigma}$ -- the $\sigma$--type 
transfer between the Ge-$sp^3$ hybrid and an O-$p$ orbital. Also the 
energies of the local unperturbed states at the Cu ($\epsilon_{0d}$) and on 
the O ($\epsilon_{0p}$) are indicated. The $nn$ superexchange is between 
Cu--0 and Cu--1, the $nnn$ superexchange between Cu--0 and Cu--2.} 
\label{figure: nn and nnn exchange c-axis} 
\end{figure} 
\begin{figure}[tbp] 
\caption[fig 8]{The original quantization axes for 90$^0$ superexchange. The 
angle $\beta$ is the deviation of the bridge angle M$_1$--L$_i$--M$_2$ ($\phi 
$) from 90$^0$. The $d_{{x^{\prime}}^2-{y^{\prime}}^2}$ orbitals on the M$_1$ 
and M$_2$ and the $p_{x^{\prime}}$ and $p_{y^{\prime}}$ orbitals on the 
ligands L$_1$ and L$_2$ are shown. The transfer matrix elements between the 
ligand $p$ and magnetic $d$ orbital is $t$. } 
\label{figure: old quantization axis} 
\end{figure} 
\begin{figure}[tbp] 
\caption[fig 9]{The new quantization axes. The axes $x^{\prime}$ and $%
y^{\prime}$ are rotated over 45$^0$ to $x$ and $y$, with respect to those 
used in fig. \ref{figure: old quantization axis}. The $\sigma$--bonding 
orbitals on the side groups S$_1$ and S$_2$ are shown. The angle $\beta$ is 
the deviation from ideal 90$^0$ geometry. The transfer integral between $%
d_{xy}$ and $p_x$ is $t$, and the transfer integral between the $p_y$ 
orbitals on the two ligands L$_1$ and L$_2$ is $W$.} 
\label{figure: new quantization axis} 
\end{figure} 
\begin{figure}[tbp] 
\caption[fig 11]{The Cu$_2$O$_2$GeSi cluster. We illustrate the local 
distortion due to Si substitution. The O which is bonded to the Si is 
shifted along the Si--O bond in the direction of Si. The bridge angle ($\phi$%
) decreases while the hinge angle ($\alpha$) increases. } 
\label{figure: Si substitution} 
\end{figure} 
\begin{figure}[tbp] 
\caption[fig 12]{Results of the band model: the $nn$ superexchange ($J_{nn}$%
: full line) and the ratio of the $nnn$ and $nn$ superexchange ($\gamma$: 
dashed line) as a function of the deviation of the bridge angle $\beta = 
\phi - 90^0$. We used the following values for the parameters: Parameters 
used: $t_{pd} = 1.0$, $W = .9$, $\epsilon_{0d} = -3.4$, $U_d = 7$, $%
V_{0\sigma} = 3.7$, $\Delta_{\sigma} = 6.65$, $J_{pd} = 0.03$, $J_{OH} = 0.6$ 
All energies are in eV. } 
\label{figure: band model bridge angle} 
\end{figure} 
\begin{figure}[tbp] 
\caption[fig 13]{Results of the band model: the $nn$ superexchange ($J_{nn}$%
: full line) and the ratio of the $nnn$ and $nn$ superexchange -- the 
frustration parameter $\gamma$: dashed line -- as a function of the hinge 
angle $\phi$. The parameters used are the same as in fig. \ref{figure: band 
model bridge angle} } 
\label{figure: band model hinge angle} 
\end{figure} 
\begin{figure}[tbp] 
\caption[fig 14]{Results of the band model: the $nn$ superexchange $J_{nn}$ 
(full line), the frustration parameter $\gamma$ (dotted line) (ratio of the $%
nnn$ and $nn$ superexchange), the derivative of $J_{nn}$ w.r.t. the bridge 
angle $\alpha$ (long dashes), and $J_{nn}$ w.r.t the hinge angle $\beta$ 
(short dashes) as a function of the Ge--O hybridization $V_{\sigma}$. The 
parameters used are the same as in fig.: \ref{figure: band model bridge 
angle} } 
\label{figure: chemical bonding factor Vsigma} 
\end{figure} 
\begin{figure}[tbp] 
\caption[fig 15]{Results of the band model: the $nn$ superexchange $J_{nn}$ 
(full line), the frustration parameter $\gamma$ (dotted line) (ratio of the $%
nnn$ and $nn$ superexchange), the derivative of $J_{nn}$ w.r.t. the bridge 
angle $\alpha$ (long dashes), and $J_{nn}$ w.r.t the hinge angle $\beta$ 
(short dashes) as a function of the energy difference between the O-$p$ and 
Ge-$sp^3$ hybrid: $d_{\sigma}$. The parameters used are the same as in fig.  
\ref{figure: band model bridge angle} } 
\label{figure: chemical bonding factor deltasigma} 
\end{figure} 
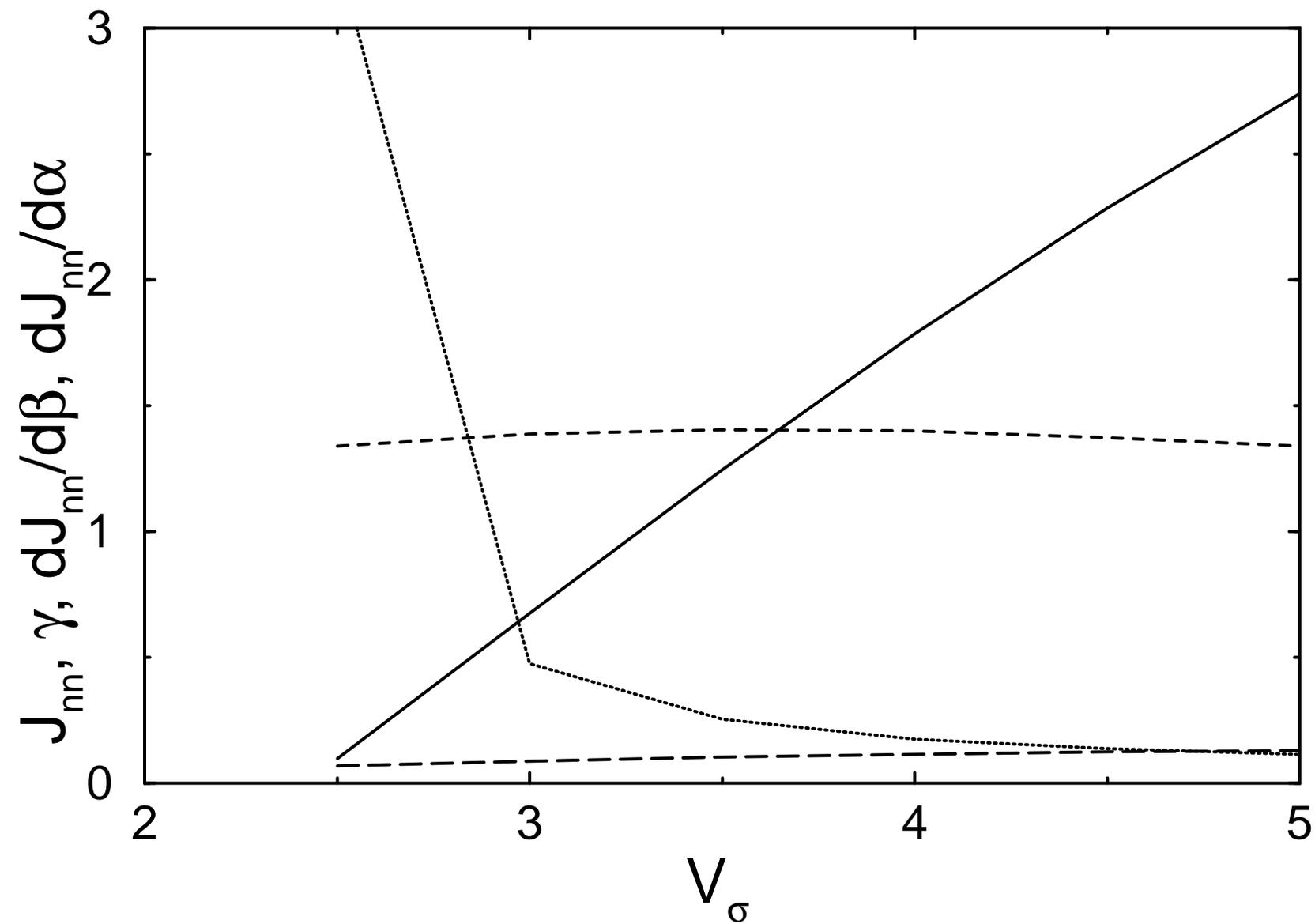
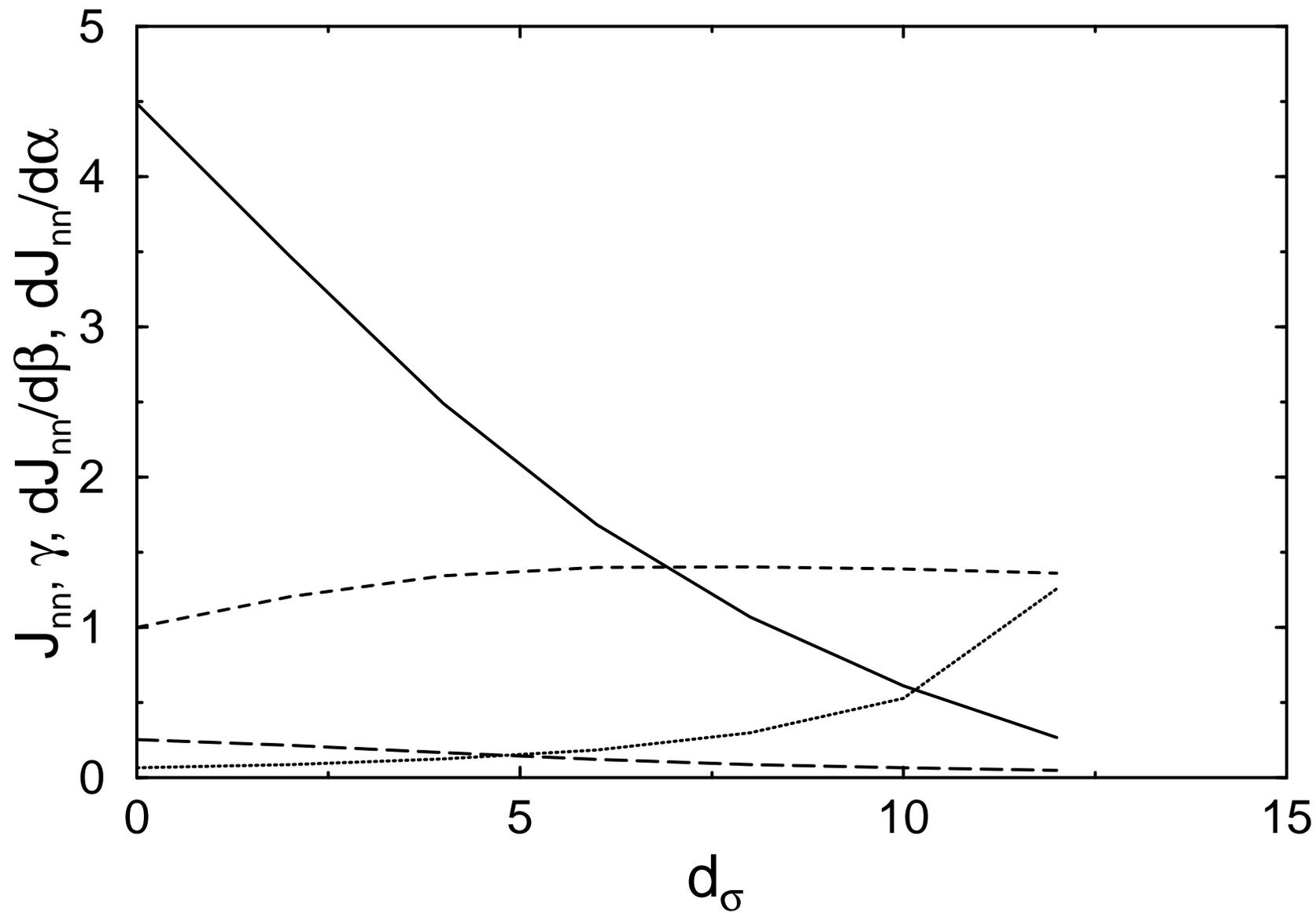
\end{document}